\documentclass[sigconf]{acmart}

\newcommand{\ignore}[1]{}

\usepackage{multirow} 
\usepackage{subcaption}

\copyrightyear{2024}
\acmYear{2024}
\setcopyright{rightsretained}
\acmConference[NSysS '24]{11th International Conference on Networking, Systems, and Security}{December 19--21, 2024}{Khulna, Bangladesh}
\acmBooktitle{11th International Conference on Networking, Systems, and Security (NSysS '24), December 19--21, 2024, Khulna, Bangladesh}
\acmDOI{10.1145/3704522.3704532}
\acmISBN{979-8-4007-1158-9/24/12}

\begin{document}

\title{Can Features for Phishing URL Detection Be Trusted Across Diverse Datasets? A Case Study with Explainable AI}


\author{Maraz Mia}
\email{mmia43@tntech.edu}
\affiliation{%
  \institution{Department of Computer Science\\ Tennessee Tech University}
  \city{Cookeville}
 \state{Tennessee}
  \country{USA}
}
\authornote{Both authors contributed equally to this research.}
\author{Darius Derakhshan}
\authornotemark[1]
\email{dxderakhsh42@tntech.edu}
\affiliation{%
  \institution{Department of Computer Science\\ Tennessee Tech University}
  \city{Cookeville}
  \state{Tennessee}
  \country{USA}
}

\ignore{
\author{Sujan Sarker}
\email{sujan@du.ac.bd}
\affiliation{%
  \institution{Robotics and Mechatronics Engineering\\ University of Dhaka}
  \city{Dhaka}
  \country{Bangladesh}
}
}

\author{Mir Mehedi A. Pritom}
\email{mpritom@tntech.edu}
\affiliation{%
  \institution{Department of Computer Science\\ Tennessee Tech University}
  \city{Cookeville}
 \state{Tennessee}
  \country{USA}
}

%\settopmatter{printacmref=false}

%\renewcommand{\shortauthors}{xyz et al.}

\begin{abstract}
Phishing has been a prevalent cyber threat that manipulates users into revealing sensitive private information through deceptive tactics, designed to masquerade as trustworthy entities. Over the years, proactively detection of phishing URLs (or websites) has been established as an widely-accepted defense approach. In literature, we often find supervised Machine Learning (ML) models with highly competitive performance for detecting phishing websites based on the extracted features from both phishing and benign (i.e., legitimate) websites. However, it is still unclear if these features or indicators are dependent on a particular dataset or they are generalized for overall phishing detection. In this paper, we delve deeper into this issue by analyzing two publicly available phishing URL datasets, where each dataset has its own set of unique and overlapping features related to URL string and website contents. We want to investigate if overlapping features are similar in nature across datasets and how does the model perform when trained on one dataset and tested on the other. We conduct practical experiments and leverage explainable AI (XAI) methods such as SHAP plots to provide insights into different features' contributions in case of phishing detection to answer our primary question, ``Can features for phishing URL detection be trusted across diverse dataset?''. Our case study experiment results show that features for phishing URL detection can often be dataset-dependent and thus may not be trusted across different datasets even though they share same set of feature behaviors. %We also recommend data merging from different sources for better generalizability in future phishing URL detection models. 

\end{abstract}


%sms phishing, smishing, phishing text, graph-based visualization, scams, smish monitoring

%\ccsdesc[300]{Security and privacy~~spam/scam messages}
%\ccsdesc[100]{Security~Smish campaign monitoring}
%\ccsdesc{Information systems~campaign monitoring system}





\maketitle

\section{Introduction}
\label{sec:intro}
Phishing attacks come in various forms, such as deceptive emails or mobile messages attached with fraudulent website URLs, all designed to trick users into revealing sensitive information or click on to malicious attachments \cite{phish_threat_graham}. Moreover, the potential abuse of generative AI and large language models may add more stress towards defenders to cope with these attacks \cite{SP2024_Phishbot_chatbot,threatGPT_Maanak2023Access,abuseGPT_ISDFS2024}. According to recent statistics, the United States alone had a total of around $300K$ phishing victims, with financial losses exceeding \$52 million %,089,159 
due to these attacks \cite{phishing_stat_M.K}. Historically, phishing website detection relied on traditional blacklisting where various publicly available blacklists like {\em PhishTank} \cite{phishtank_portal} and other private blacklists are leveraged. \ignore{However, with the advance of machine learning and deep learning models, these methods are leveraged in literature as well.} \ignore{These techniques use various features extracted from URLs and website content to train ML models in a supervised manner where phishing and benign websites are used to compare the detection performance.} While these black-box detection models may achieve high accuracy, they lack transparency and explainability. Due to this shortcoming, black-box models hinder trust and adoption in practice. Moreover, the dynamic nature of phishing website data involves concept drift \cite{menon2021concept}, which describes a situation where the relationship between the input data and the target variable varies over time in an online supervised learning environment. Although detecting the concept drift and retraining the model with newly extracted features \cite{8455975} can partially resolve the issue, the overall feature importance in different deployment scenarios with different schemes of features can still be varied and not generalized.%, particularly in high-stakes domains like cybersecurity.

To bridge this gap, SHAP (SHapley Additive exPlanations), a popular explainable AI (XAI) method, can be used to interpret the individual (i.e., local explanation) and overall model predictions, which %To contextualize, SHAP is a unified measure of feature importance that allocates each feature an importance value for a particular prediction \cite{NIPS2017_7062_shap}, and it is 
can aid in the decision-making process \cite{NIPS2017_7062_shap}. %\footnote{this SHAP intro can be done in the intrpoduction section....where you mentioned SHAP in paragraph 2 of intro} 
\ignore{This approach allows us to understand the importance of various features in detecting phishing URLs or websites.} %\footnote{the motivation or background on Phishing Dataset related issues is not highlighted until this point, which should be here before we propose our statement for this paper!}
In this paper, we propose to leverage XAI approaches to understand the generalization of phishing URL detection features across datasets. We incorporate XAI as a means to provide insight into the model’s decision-making process, shedding light on features which are more impactful in the classification of an instance as {\em phishing} versus {\em benign}. Our primary objective is to answer the following question- \textit{``Can features' importance for phishing URL detection be trusted across diverse datasets?''}. By answering the question, we want to know if certain set of features are ubiquitous for phishing URLs detection, or if the features are closely tied to a specific dataset.  % This approach not only enhances the robustness of phishing detection explanations but also promotes understanding, transparency, and trust in machine learning models.

In addition, we also evaluate the performance of different ML models on multiple datasets to select the best ML model for generating SHAP explanations. Furthermore, we create various experiment scenarios where training and testing portion of one dataset is used with another dataset. This is particularly beneficial when common overlapping features are present in multiple datasets, and training and testing with different datasets can provide insights on their generalizability. %Moreover, in real-world one dataset may not capture various phishing URLs and merging them  as it assists in assessing the effectiveness of similar features in new datasets and also in real-world scenarios where diverse type of phishing URLs may not be captured in a single dataset. 
We hereby hypothesize that the claimed accuracy of any particular ML model achieved by the researchers on a specific dataset, may get declined while the test environment changes or new data appears. If this is true, then we have got our answer for the primary question and need to be cautious about phishing detection results. %Thus, the incorporation of transferability of features of a model that is trained on one dataset, but tested  when trained  not only strengthens the explanation of underlying patterns but also provides generalized prediction adaptability across different datasets.
\ignore{This experimental case study paper should provide some useful insights for researchers and practitioners in phishing detection domain as we often find highly accurate detection rates and effective feature lists in literature of machine learning based detection, which may be specific to certain dataset and may not generalize when diverse dataset are introduced.} 
%to show results of higher detection witcontributing to the ongoing efforts to make the internet a safer place from a pervasive phishing threat.

%\vspace{-2mm}
%\subsection{Research Questions}
%vspace{-0.5em}

To guide our experiments in this paper, we are driven by the following \textbf{three research questions (RQs)}. % fIn this paper, we address the following \textbf{four} \textbf{research questions (RQs)} through different experimental case study scenarios.  %The methodology used in the experimental research for the detection and explainability of phishing websites/URLs is centered around answering the following research questions.

\noindent\textbf{RQ1:} What are the top impactful overlapping features and their impact distribution for a specific prediction outcome across multiple datasets for phishing detection?

 %How can Explainable AI be used to determine which features are most influential in detecting phishing websites, and do these features vary significantly between datasets?

\noindent\textbf{RQ2:} When multiple datasets share overlapping features, how do a ML-based phishing detection model perform, when trained on one dataset and tested on another dataset? Does it improve the detection performance if both datasets are merged for training? %

\noindent\textbf{RQ3:} Are overall features' contribution ranks for the shared overlapping phishing URL features showing a similar contribution order in different datasets? %Does it provide any new insights? % ubiquitous or dataset dependent?

%\noindent\textbf{RQ4:} Can explainable AI help provide newer insights with features' contribution trends and rank orders in different dataset scenario? %\footnote{I think in our current insights we are showing there is a change ....but why is that change, we need to provide those insights in our analysis very clearly} 
% from another dataset, given theCan a model trained to detect phishing attempts on one dataset effectively predict phishing attempts on a different dataset?
% and what challenges does it face when transferring the model between these datasets?

In summary, motivated by these above research questions we make the following major contributions in this paper: %MotivatedIn this paper, we make the following key contributions:

\begin{itemize}
    \item Analyze overlapping features from multiple phishing URL datasets consisting more than $108K$ unique URLs.
    %\item Various machine learning models have been implemented to determine which can most accurately classify phishing URLs, such as Linear Regression (LR), Decision Tree (DT), Random Forest (RF), Naive Bayes (NB), Gradient Boosting Machine (GBM), XGBoost (XGB), Explainable Boosting Machine (EBM), and Support Vector Classifier (SVM).
    \item Answer the RQs with experimental evidences if features for phishing URL detection can be generalized across datasets where training and testing of ML models are conducted on different datasets.  %Train a model on a specific dataset and testing on other datasets is utilized to evaluate the predictive power of common features across datasets, even when the model has no prior information about the test portion of the dataset.
    \item Use popular XAI SHAP module to provide new insights and find deviations in features' contribution behaviors for phishing detection when multiple datasets are involved.
   % \item Generating explanation plots on feature contribution which provide valuable insights on different cases to check whether the feature contribution remained monolithic even if we use the same feature list from different datasets.
\end{itemize}

The rest of the paper is organized as follows: Section \ref{sec:rel_work} discusses related works on AI based phishing detection. %Section \ref{sec:method} overviews the main research questions addressed and overview of research methodology. 
Section \ref{sec:case_study} presents the methodology and results with data-driven insights from the experiments. Section \ref{sec:discuss} discuss the current state and limitations in the present study while Section \ref{sec:conclusion} concludes the paper.

%\vspace{-2mm}
\section{RELATED WORKS}
\label{sec:rel_work}
% \footnote{need to shorten the related work section}
ML and Deep Learning (DL) based phishing detection research has seen significant progress with the advancement of the Artificial Intelligence. In literature, there are number of studies proposed to incorporate ML for phishing URL, malicious domain or website detection and supporting law-enforcement take-down decisions based on URL features and webpage contents \cite{b2,b3,b4,zamir2020phishing, pritom2020_covid_malwebsites, pritom2022_cns_website_lawenforcement,abdelhamid2017phishing,codaspy_2013_xu_cross_layer_malwebsite_detection,jain2018phish,gupta2021novel,comar_euro_sp2020_maroofi,denis_mal_domain_2020}. However, a common limitation across these studies is the lack of interpretability of the underlying decision-making processes of the models used, often referred to as the ``black box'' problem. Furthermore, there is a scarcity of research that compares the performance and feature importance across diverse datasets. Sarasjati \textit{et al.} \cite{b7} and Preeti \textit{et al.} \cite{b8} presented comparison of various ML models across multiple phishing website datasets, and the usage of multiple datasets with varying class labels allows for a more robust analysis. These studies may provide insights about the best available ML models with a given set of features, but they can not provide any insights on the interpretation or explainability of the models. Next, Rugangazi \textit{et al.} \cite{b9} proposed an automated phishing detection strategy that picks important features using the global feature importance method to achieve high accuracy but it is limited to a single dataset and does not discuss model interpretability to assess the features generalizability for phishing detection. %The researcher also introduced different techniques and strategies to identify phishing websites based on several modes of features. 
In another study, Ali \textit{et al.} \cite{aljofey2020_cnn_based_phishing} proposed to rely on URL character sequences using character-level convolutional neural network (CNN) while Tao \textit{et al.} \cite{feng2020visualizing} proposed character-level recurrent neural network (RNN) for phishing detection. However, these models mostly lack transparency for detection of phishing versus benign URLs, which may be targeted by adversaries by tweaking domain names to avoid detection. Moreover, Youness \textit{et al.} \cite{mourtaji2021hybrid_cnn} showed CNN outperforming %the best performance in phishing detection using 37 features from six different cases which includes 
blacklisting, lexical, content-based, and visual \& behavioral similarity methods. In literature, we also find articles where natural language processing (NLP) is used with ML to effectively detect phishing attempts \cite{sahingoz2019_ML_for_phishing}. 

%\ignore{Likewise, the model or ML algorithms for phishing detection, and the feature selection process is vital in automated phishing website detection.} 

In phishing research, there are a number of features explored in literature, but among them lexical features are the ones those are very easily available and used mostly \cite{verma2017s_malicious_url, rao2020catchphish_url}. Other than lexical features, we observe URL statistics, HTML code, webpage javascript, webpage text, website external links, website structure, domain ranking, SSL certificates, TLD reputation, WHOIS data, DNS, and passive DNS based features \cite{das2024modeling_hybrid_features,opara2024look_url_html, aljofey2022effective_html_url}. We also find pre-trained transformer-based models such as BERT for feature extraction from websites \cite{elsadig2022intelligent_bert}, which suffers from dataset-dependency biases. %Guptta \textit{et al.} \cite{} implemented URL lexical and hyperlink-based features and showed that phishing websites normally have more external website links other than the base domain. 

Furthermore, researchers have proposed benchmark datasets for phishing detection with combination of four categories of features based on URL, website's contents (HTML and Javascipts), external third-party (WHOIS, Google, OpenPageRank) or visual similarity using common sources (e.g., PhishTank, OpenPhish, Alexa, PhishMonger) \cite{chiew2018building,hannousse2021towards,zeng2020diverse,el2020depth}. \ignore{With that intuition, Hannousse \textit{et al.} created a dataset with $87$ features which exclude the visual features \cite{hannousse2021towards}. In another study, PhishBench \cite{zeng2020diverse,el2020depth} has provided a URL and website based dataset with $83$ features, where the visual features are not included as they are not consistent across datasets.} In our case studies, we have analyzed the features that are common within these benchmark datasets. %So, eventually within the two datasets further investigated in this paper have mostly overlapping features with the existing benchmark datasets along with additional diverse features. 
%However, in this study our goal is to find common features between datasets that can be used for between dataset analysis.  % , we have found along with additional features those are observed in the two datasets further investigated in this paper. %there have some certain types of website sources and some limited feature types. However, we have chosen 2 different datasets other than the highlighted and claimed benchmark datasets because these two datasets contains more diverse features (first dataset has 111 and second dataset has 87) curated from the mentioned sources (also excluding the visual based features).}\footnote{must address}
In summary, all the existing approaches either lack transparency (i.e., interoperability) or generalizability (i.e., reproducing results with another dataset context). %In this paper, we are showing with experimental results how the same feature sets may have different impacts on different datasets. \ignore{Thus, if phishing detection is shown on a smaller dataset with a list of effective features, it may not be generalized and trusted across datasets.} 
In this paper, we showcase how XAI can bring transparency of detection models as well as how they can be leveraged to test the generalizability of models across datasets. 
% outcomes in different datasets. 

%\section{Research Questions and Case Study}
%\label{sec:method}
%\input{methodology.tex}

%\vspace{-1.5mm}
\section{Experimental Case Studies and Insights}
\label{sec:case_study}
%\vspace{-0.2em}

%\subsection{Overview of Methods}
%\vspace{-0.35em}
\begin{figure*}[!t]
    \centering
    \includegraphics[scale=0.80, width=0.9\textwidth]{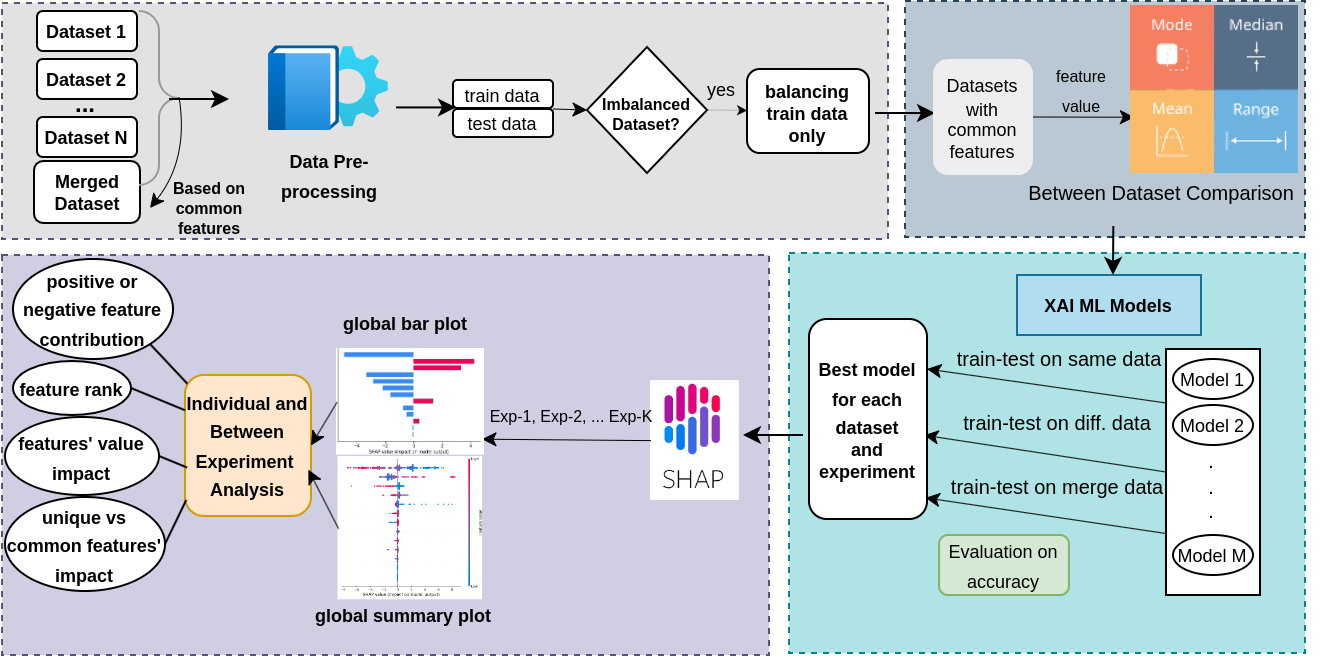}
    \caption{Overview methods for evaluating phishing features' generalizability}
    \label{fig:methodology}
\end{figure*}

To address the RQs, we propose the methodology highlighted in Fig. \ref{fig:methodology}, which has the following five components--
(i) Datasets collection and pre-processing; (ii) Feature analysis; (iii) Train and test XAI models; (iv) Evaluate model performances in various dataset-based experiment scenarios; (v) Generate insights from XAI outputs.

%\vspace{-0.25cm}

\subsection{Data Collection and Data Pre-processing}
%In this study, we choose two datasets containing different types of features for phishing and legitimate websites.

\subsubsection{Dataset-1 ($D_1$)} This dataset is collected from Vrbančič \textit{et al.} \cite{b10} and consists of $88,647$ instances with $58,000$ being benign and $30,647$ being phishing URLs. Additionally, there are $111$ features in this dataset taken from URL attributes and web contents. The dataset annotators considered different types of feature columns based on the whole URL, domain name, URL directory, URL file name, URL parameters, resolving URL, and third-party services. 

%\vspace{-1mm}
\subsubsection{Dataset-2 ($D_2$)} This dataset is collected from Kaggle \cite{b11} that consists of $19,431$ row entries with $9,716$ being benign and $9,715$ being phishing samples. There are $85$ total features in this dataset including features extracted from the URL, the HTML content, and the web domain.

%\vspace{-1mm}
\subsubsection{Data Pre-processing} By analyzing the above two datasets, $D_1$ and $D_2$, we find $20$ common features between them (while it is important to note that some common features have different names given by the annotators in these datasets, which is addressed by manually looking at the feature definitions of both datasets).

\noindent \textbf{Addressing missing values} If there is a null value in any rows for a particular feature (i.e., a column in the dataset), we fill it with the median value of that corresponding feature column. 

\noindent \textbf{Removal of features} We have removed the columns that have no insights and only provided a single constant value in all rows. %(e.g., {\color{red}column x}\footnote{what is column x? any specific definition of x?} has a value of 1 for all rows). 
The preprocessing steps resulted in dataset $D_1$ being reduced to $98$ features, $D_2$ being reduced to $79$ usable features.
% , while $D_{merge}$, the amalgamation of $D_1$ and $D_2$ still with $20$ features. %\footnote{Please check, if these column counts include the target column of phishng=1 and benign=0? if yes, then we need to state 1 less columns for features {\color{blue}All 20 columns are given in the common feature table 2, no label column is included.}} %To minimize confusion, we have renamed all the common feature columns into one singular and similar naming convention for all the datasets. 
%Finally, target variables were separated from the features and transformed into integers (0 for benign, 1 for phishing) to facilitate the training of various classical machine learning models.

\begin{table}[!h]
\centering
%\vspace{-3mm}
\caption{Final train-test data distributions (SMOTE applied on $D_1$ training portion only for data balancing)}
\resizebox{0.35\textwidth}{!}{
\begin{tabular}{|c|cccc|}
  \hline
  \multirow{2}{*}{\textbf{Dataset}} & \multicolumn{2}{c}{\textbf{Train}} & \multicolumn{2}{c|}{\textbf{Test}} \\
  \cline{2-5}
    & Phish & Benign & Phish & Benign\\
  \hline
  $D_1$ & 40,614 & 40,614 & 9,209 & 17,386 \\
  % \hline
  $D_2$ & 6,770 & 6,831 & 2,945 & 2,885 \\
  % \hline
  $D_{merge}$ & 13,570 & 13,631 & 12,154 & 20,271 \\
  \hline
\end{tabular}
%\vspace{-3mm}
\label{tab:dataset_cnt}
}
\end{table}

\subsubsection{Correcting Data Imbalance and Optimizing Data Splitting} Dataset $D_2$ is already balanced for both classes, while we observe imbalance in $D_1$, with $88,647$ {\em benign} and $30,647$ {\em phishing} instances (ratio of $2.9:1$). %as seen in the raw data from Dataset-1 and the combined Dataset-1+2, there is an apparent imbalance in the distribution between benign and phishing labels. 
This imbalance can lead to biased models, as they tend to more accurate in predicting the majority class, leading to poor generalization performance when predicting minority class. \ignore{\cite{dataset_imbalance_Leevy2018}} To address this issue, we adopt the Synthetic Minority Over-sampling Technique (SMOTE) \cite{b5}, which is a popular oversampling method that generates synthetic instances of the minority class by interpolating between existing minority instances. We apply SMOTE only on the training portion of dataset $D_1$. This can produce more comparative model outcomes with the already balanced $D_2$. In this paper, we adopt a $70:30$ split ratio for the train-test data splits. %ratio cases, we have found better results for . %distributed for a more direct comparison. 
%We need to apply SMOTE only in the training portion of the dataset, else the test result from model prediction won't be reflected on the original dataset because the sampled dataset will contain many redundant entries, and if we take the test data from that, clearly we will get higher accuracy but still biased. 
% Also, we have used random undersampling or mixed of both SMOTE and random undersampling. We will use one specific sampling method based on the model's best performance. 
After correcting class imbalance, for some experiment scenarios, we have merged the two datasets $D_1$ and $D_2$ considering the $20$ common features (defined in table \ref{table:feature_def}) and created a third dataset, $D_{merge} = D_{merge(train)} + D_{merge(test)}$.
% resulting into $108,078$ unique instances where $67,716$ being benign and $40,362$ being phishing samples in this combined dataset.
To maintain the class balance and to generate fair explanation from SHAP, in $D_{merge(train)}$, we have taken 6,800 random instances for each groups (phishing and benign) in $D_{1}$'s training portion and merged them with the $D_2$'s training portion to get a total of $13,570$ phishing instances and $13,631$ benign instances. The test data for the merged dataset $D_{merge}$ is the direct concatenation of the test data for $D_1$ and $D_2$. Table \ref{tab:dataset_cnt} depicts the distribution of all datasets $D_1$, $D_2$, and $D_{merge}$. 
\subsection{Feature Analysis}
%\vspace{-0.45em}
In both datasets $D_1$ and $D_2$, we have found that the majority of the features are extracted from the URL string (i.e., lexical features). Moreover, dataset $D_1$ does not have any HTML and JavaScript features while those are present in $D_2$ (i.e., existence of login form, iframe, favicon-based external links, and click event). So, between these two datasets, there are differences in the feature list if we consider all features. However, we want to know the impact of unique features ($F_u$) and the common features ($F_c$) as listed in Table \ref{table:feature_def} on the prediction models. That is why we consider the full feature list ($F_u \cup F_c$) along with only the common ones ($F_c$) to evaluate prediction model performance. %which we gained after the data pre-processing step.%\footnote{need to discuss for clarification {\color{blue}done}}

\begin{table}[!t]
    \centering
    \caption{Feature definitions for common features ($|F_c|=20$)}
    \resizebox{0.48\textwidth}{!}{
    \begin{tabular}{|p{0.25 cm} c c|}
    \hline
    \textbf{ID} & \textbf{Feature Name} & \textbf{Feature Definition} \\
    \hline
    $f_{1}$ & {\tt qty\_dot\_url} & Number of dot characters `.' in URL \\
    % \hline
    $f_{2}$ & {\tt qty\_equal\_url} & Number of '=' character in URL \\
    % \hline
    $f_{3}$ & {\tt domain\_length} & Length of the domain name string \\
    % \hline
   $f_{4}$ & {\tt url\_google\_index} & If the URL is indexed by Google \\
    % \hline
    $f_{5}$ & {\tt qty\_dollar\_url} & Number of '\$' character in URL \\
    % \hline
    $f_{6}$ & {\tt qty\_slash\_url} & Number of '/' character in URL \\
    % \hline
    $f_{7}$ & {\tt qty\_redirects} & Number of redirects for landing page \\
    % \hline
    $f_{8}$ & {\tt url\_shortened} & If the URL is shortened %(e.g., using services like Bit.ly) 
    \\
    % \hline
    $f_{9}$ & {\tt tld\_present \_params} & If TLD present in the parameters of URL \\
    % \hline
    $f_{10}$ & {\tt qty\_comma\_url} & Number of ',' character in URL \\
    % \hline
    $f_{11}$ & {\tt qty\_hyphen\_url} & Number of '-' character in URL \\
    % \hline
    $f_{12}$ & {\tt qty\_underline\_url} & Number of '\_' character in URL\\
    % \hline
    $f_{13}$ & {\tt length\_url} & Length of entire URL \\
    % \hline
    $f_{14}$ & {\tt qty\_percent\_url} & Number of '\%' character in URL \\
    % \hline
    $f_{15}$ & {\tt qty\_asterisk\_url} & Number of '*' character  in URL \\
    % \hline
    $f_{16}$ & {\tt qty\_questionmark} \tt \_url & Number of '?' characters in URL \\
    % \hline
    $f_{17}$ & {\tt qty\_tilde\_url} & Number of '$\sim$' character in URL \\
    % \hline
    $f_{18}$ & {\tt qty\_at\_url} & Number of '@' character in URL\\
    % \hline
    $f_{19}$ & {\tt domain\_in\_ip} & If the domain is an IP address \\
    % \hline
    $f_{20}$ & {\tt qty\_and\_url} & Number of '\&' character in URL\\
    \hline
    \end{tabular}
    \label{table:feature_def}
    }
\end{table}

\begin{figure}[!t]
    \centering
    \includegraphics[width=0.85\columnwidth, height=0.69\linewidth]{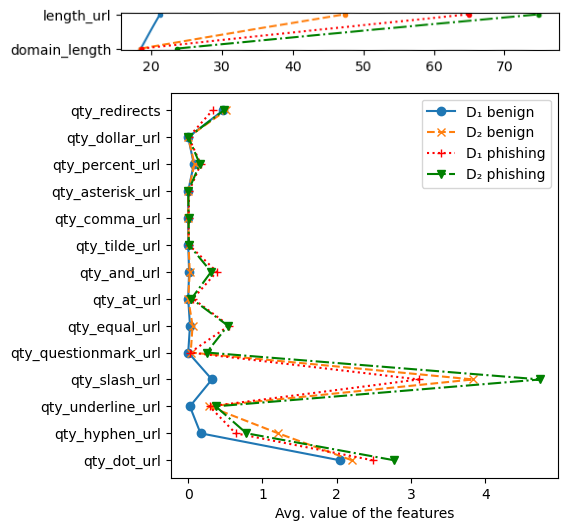}
    \caption{Mean values for the common {\em numerical} features}
    \label{fig:feat_avg_phish}
\end{figure}

\begin{figure}[!h]
    \centering
    \includegraphics[width=0.85\columnwidth, height=0.65\linewidth]{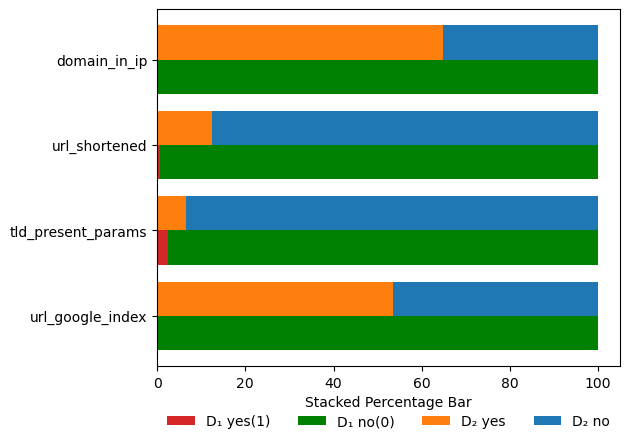}
    \caption{Percentages for the common {\em binary} features}
    \label{fig:feat_avg_non_phish}
    %\vspace{-1mm}
\end{figure}

\ignore{We also investigate to understand if the $20$ common features (see Table \ref{table:feature_def}) found in both datasets are important ones in general for phishing detection or not as those are used further to conduct the between dataset experiment.} %We have described the definitions of $20$ common features in .
We also provide the basic statistics (i.e., mean values) comparison plot for both datasets in terms of phishing and benign URLs as shown in Fig. \ref{fig:feat_avg_phish} and Fig. \ref{fig:feat_avg_non_phish}. This analysis is important, because if there exists innate differences in the dataset feature values, then it indicates an obvious difference in the explanations as well. % will also be different. From the figures \ref{fig:feat_avg_phish} and 
Now, Fig. \ref{fig:feat_avg_phish} shows that the mean values for the numerical features are very similar in both the datasets. However, in Fig. \ref{fig:feat_avg_non_phish}, the percentages of binary features' values for {\tt doimain\_in\_ip} and {\tt url\_google\_index} features are highly deviating between the datasets while the other binary features have very similar distribution. Additionally, we have also checked other statistical values such as min, max, median, and standard deviation for the features and observe a very similar values across the datasets.  %major differences across both datasets.%The common features were put into a dictionary so that despite the varied naming conventions, they would able to be used for both datasets.
%\footnote{where is the analysis? where is the selection part here? if we select all features, then ``feature selection'' for section title is misleading.}

\subsection{Train and Test with Supervised Models}
Next, we experiment with various machine learning models that can be further incorporated with SHAP XAI module. %as we keep to gain feature based insights from the model outcomes. %This case study implements a variety of machine learning models, each with its unique strengths and characteristics, to ascertain the model that most accurately classifies phishing websites. The models used are:
We train and test with different dataset portions in various experiment scenarios leveraging models such as {\em Logistic Regression} (LR), {\em Decision Tree} (DT), {\em Random Forest} (RF), {\em Naive Bayes} (NB), {\em Gradient Boosting Machine} (GBM), {\em XGBoost} (XGB), {\em Explainable Boosting Machine} (EBM) and {\em Support Vector Machine} (SVM). The performances of these ML models are evaluated using the standard evaluation metrics- \textit{Accuracy}, \textit{Precision}, \textit{Recall}, and \textit{F1 Score}. Among both the $D_1$ and $D_2$ datasets, {\em XGBoost} (XGB) performs the best for the detection of phishing URLs as shown in Table \ref{tab:eval_models} with an accuracy of $97.1\%$ in dataset $D_1$ and $99\%$ in dataset $D_2$ when considered all features. The XGB model is followed by {\em Random Forest} with an accuracy of $97\%$ for $D_1$ and  $98.6\%$ for $D_2$. Then, in order from highest accuracy to least accuracy we observe {\em Explainable Boosting Machine} (EBM), {\em Decision Tree}, {\em Gradient Boosting Machine}, {\em Logistic Regression}, {\em Naive Bayes}, and {\em Support Vector Machine} models. %From the above models, clearly XGBoost outperforms all the other models. %and also it is suitable to use with SHAP explanation.  %top-performing model, it is nearly identical to Random Forest and more suitable for use with SHAP. Unlike Random Forest, which outputs probabilities for each class, XGBoost outputs one probability, making it more straightforward to obtain SHAP values. 
Thus, we use the {\em XGBoost} model to generate SHAP explanation plots for the various experiment scenarios in finding the feature's contributions.

% Accuracy is the ratio of correctly predicted instances to the total instances.
% \begin{align} 
% &\mathit{Accuracy} = \frac{TP+TN}{TP+TN+FP+FN}
% \end{align}

% Precision is the ratio of correctly predicted positive instances to the total predicted positive instances.
% \begin{align}
% &\mathit{Precision} = \frac{TP}{TP+FP}
% \end{align}

% Recall (also known as sensitivity) is the ratio of correctly predicted positive instances to all actual positive instances.
% \begin{align}
% &\mathit{Recall} = \frac{TP}{TP+FN}
% \end{align}

% F1 Score is the harmonic mean of precision and recall, providing a balance between the two metrics.
% \begin{align}
% &F1 = \frac{2*\mathit{Precision}*\mathit{Recall}}{\mathit{Precision}+\mathit{Recall}}
% \end{align}

% \begin{figure*}[!t]
%     \centering
%     \includegraphics[width=\textwidth]{Figure/df1_mdl_cmp.png}
%      \caption{Comparison of ML models on dataset $D_1$}
%     \label{fig:df1cmp}
% \end{figure*}

% \begin{figure*}[!h]
%     \centering
%     \includegraphics[width=\textwidth]{Figure/df2_mdl_cmp.png}
%     %\vspace{-0.8em}
%     \caption{Comparison of ML models on dataset $D_2$}
%     \label{fig:df2cmp}
% \end{figure*}

% \begin{figure*}[!htbp]
%     \centering
%     \includegraphics[width=\textwidth]{Figure/dfmer_mdl_cmp.png}
%      \caption{Comparison of ML Models of Dataset-1+2}
%     \label{fig:df3cmp}
% \end{figure*}

%\vspace{-1.5mm}
\subsection{Model Evaluation in Different Experiments}
\label{sec:evalaution_experiment_setting}
To strengthen our hypothesis \ignore{ \footnote{did we define a hypothesis earlier? in which section? {\color{blue} Sir, this is mentioned in the introduction section.}}} %\footnote{which hypothesis exactly? not clear to me.. {\color{blue} Sir, it was present in the previous version of the paper in the intro section, I have re-added the hypothesis. Like the claimed accuracy of the model by the researcher may get changed if the test environment vary than the actual dataset.}} 
on declining model accuracy in multiple dataset scenarios, we conduct several experiments with different training and testing data portions. The experiment scenarios include finding features' contributions when training and testing the model on the same dataset with all features, %($Train-Test_{Cat_1}$), 
training and testing the model on the same dataset with only common features, %($Train-Test_{Cat_2}$) 
and training the model on one dataset but testing on a different dataset, and finally training and testing on a merged dataset. 
% with common features. %($Train-Test_{Cat_3}$). 
We use the {\em XGB} model in this experiment scenarios to understand the features' behavior. Table \ref{tab:eval_exp} shows that using all features in a train-test scenario on the same dataset %or $Train-Test_{Cat_1}$ 
(i.e., Exp-1 and Exp-2) provides us with the best accuracy. Also, only using the 20 common features and conducting the train-test on the same dataset % in $Train-Test_{Cat_2}$ 
(i.e., Exp-3, Exp-4) gives us reasonably good accuracy results. However, when we apply train-test scenarios from different datasets, %$Train-Test_{Cat_3}$ 
the accuracy drops drastically to $51\%$ when model trained on $D_1$ and tested on dataset $D_2$, and $59\%$ when trained on $D_2$ and tested on $D_1$. %\footnote{please put the other metrics like precision,recall for this result {\color{blue}done}}. %in Exp-5 and 59\% in Exp-6, respectively. 
In contrast, with the merged dataset $D_{merge}$ in Exp-7, the accuracy is drastically increased to 91\%, which can be implemented to get more generalized results.

\begin{table}[!t]
\centering
\caption{Comparison of different ML models on both datasets (in percentage)}
%\vspace{-1mm}
\resizebox{0.48\textwidth}{!}{
\begin{tabular}{|c|p{0.8cm}*{4}{p{0.3cm}}*{4}{p{0.48cm}}|}
\hline
\textbf{Dataset} & \textbf{Metrics} & \textbf{LR} & \textbf{DT} & \textbf{RF} & \textbf{NB} & \textbf{GBM} & \textbf{XGB} & \textbf{EBM} & \textbf{SVM}\\
\hline
% \multirow{1}{*}
 & Acc. & 90.4 & 95.5 & 97.0 & 86.0 & 94.9 & 97.1 & 96.8 & 71.6\\ 
  \cline{2-10}
  $D_1$ & Prec. & 86.0 & 93.1  & 94.5 & 88.1 & 90.5 & 94.9 & 94.8 & 55.8\\
   \cline{2-10}
 & Recall & 86.3 & 94.0 & 96.9 & 68.9 & 95.4 & 96.7 & 96.2 & 86.3\\
  \cline{2-10}
 & F-1 & 86.2 & 93.6 & 95.7 & 77.3 & 92.9 & 95.8 & 95.5 & 67.7\\
 \hline

 & Acc. & 80.0 & 97.2 & 98.6 & 74.1 & 96.2 & 99.0 & 98.2 & 60.2\\ 
  \cline{2-10}
  $D_2$ & Prec. & 81.4 & 97.0 & 98.5 & 71.2 & 96.4 & 99.1 & 98.3 & 56.8\\
   \cline{2-10}
 & Recall & 78.3 & 97.5 & 98.7 & 81.8 & 96.2 & 98.8 & 98.1 & 88.1\\
  \cline{2-10}
 & F-1 & 79.8 & 97.2 & 98.6 & 76.2 & 96.3 & 99.0 & 98.2 & 69.1\\
 \hline
 
\end{tabular}
\label{tab:eval_models}
}
\end{table}

\begin{table}[h!]
\centering
\caption{Performances of the {\em XGB} model using different train-test dataset experiment scenarios}
%\vspace{-1.5mm}
\resizebox{0.48\textwidth}{!}{
\begin{tabular}{|c| c c p{0.65 cm}p{0.65 cm}p{0.7 cm}p{0.7 cm}|}
\hline
\textbf{Features} & \textbf{ID} & \textbf{Experiment} & \textbf{Acc.} & \textbf{Prec.} & \textbf{Rec.} & \textbf{F1}\\ \hline
% \multirow{1}{*}
All (98) & Exp-1 & train on $D_1$, test on $D_1$ & 97.1 & 94.9 & 96.7 & 95.8\\ 
  All (79) & Exp-2 & train on $D_2$, test on $D_2$ & 99.0  & 99.1 & 98.8 & 99.0\\  \hline
 & Exp-3 & train on $D_1$, test on $D_1$ & 92.0  & 92.0 & 92.0 & 92.0\\  
 & Exp-4 & train on $D_2$, test on $D_2$ & 93.0  & 93.0 & 93.0 & 93.0\\   
 {Common} & Exp-5 & train on $D_1$, test on $D_2$ & 51.0  & 59.0 & 51.0 & 36.0\\  
 {(20)} & Exp-6 & train on $D_2$, test on $D_1$ & 59.0 & 70.0 & 58.0 & 50.0 \\   
 & Exp-7.1 & train on $D_{merged}$, test on $D_{merged}$ & 91.0  & 92.0 & 91.0 & 91.0\\  
 & Exp-7.2 & train on $D_{merged}$, test on $D_{1}$ & 91.0  & 91.0 & 91.0 & 91.0\\
 & Exp-7.3 & train on $D_{merged}$, test on $D_{2}$ & 92.0  & 92.0 & 92.0 & 92.0\\  \hline
\end{tabular}
\label{tab:eval_exp}
}
\end{table}
%\vspace{-.5em}

% \begin{table}[h!]
% \centering
% \caption{Performances of the {\em XGB} model using different train-test dataset experiment scenarios}
% \begin{tabular}{|c|p{1 cm}p{0.65 cm}p{0.65 cm}p{0.7 cm}p{0.7 cm}|}
% \hline
% \textbf{Features} & \textbf{ID} & \textbf{Acc.} & \textbf{Precs.} & \textbf{Recall} & \textbf{F1-Score}\\ \hline
% % \multirow{1}{*}
% All (98)  & Exp-1 & 0.97 & 0.95 & 0.97 & 0.96\\ 
%   All (79) & Exp-2 & 0.99  & 0.99 & 0.99 & 0.99\\  \hline
%  & Exp-3 &  0.92  & 0.92 & 0.92 & 0.92\\  
%  & Exp-4 &  0.93  & 0.93 & 0.93 & 0.93\\   
%  {Common (20)} & Exp-5 & 0.51  & 0.59 & 0.51 & 0.36\\  
%  & Exp-6 &  0.59 & 0.70 & 0.58 & 0.50 \\   
%  & Exp-7.1 & 0.91  & 0.92 & 0.91 & 0.91\\  
%  & Exp-7.2 &  0.91  & 0.91 & 0.91 & 0.91\\
%  & Exp-7.3 &  0.92  & 0.92 & 0.92 & 0.92\\  \hline
% \end{tabular}
% \label{tab:eval_exp}
% \end{table}

\subsection{Insights with XAI Using SHAP Plots}
In Explainable AI, feature importance can be classified into two categories: local and global. Local relevance refers to each feature's contribution to the prediction of a single instance, whereas global importance assesses each feature's overall impact on all instances. \ignore{While both categories of importance offer useful information, the following experimental research focuses on global importance to understand the feature trends in different datasets' contexts.} Phishing attempts are ubiquitous and diversified, therefore global importance is necessary to discover features that regularly influence the model's predictions, providing a more comprehensive view of feature significance. \ignore{So, the SHAP Global Bar Plot will be an effective tool for showing the global importance trends of features for a particular model.}Here we apply the SHAP \emph{TreeExplainer} object on the XGB model in all 7 experimental scenarios. Then, we have taken an equal number of phishing and legitimate instances from the test data and fed that to the explanation module to generate the corresponding SHAP values and plots. Each bar in the bar plot corresponds to a feature used in the phishing website detection model, where the length of the bar represents the mean absolute SHAP value for that feature - a measure of the impact of the feature on the model's output. A longer bar indicates a higher average impact, meaning that the feature strongly influences the model's predictions. This bar plot is also color-coded where the red bar represents the overall positive contribution and the blue bar represents a negative contribution of a specific feature. \ignore{In general, the red bar features are key indicator features for phishing (i.e., positive) and the blue ones are for legitimate (i.e., negative) predictions.} The summary plot provides us the overall picture of how each feature's actual value range (high, medium, or low) impacts the prediction for categorical target variable (i.e., {\em phishing} and {\em benign} in our case). Below, we discuss our findings of the features' importance for each of the 7 experiment scenarios.

\begin{figure}[!t]
    \centering
    \includegraphics[width=1\columnwidth, height=1.25\linewidth]{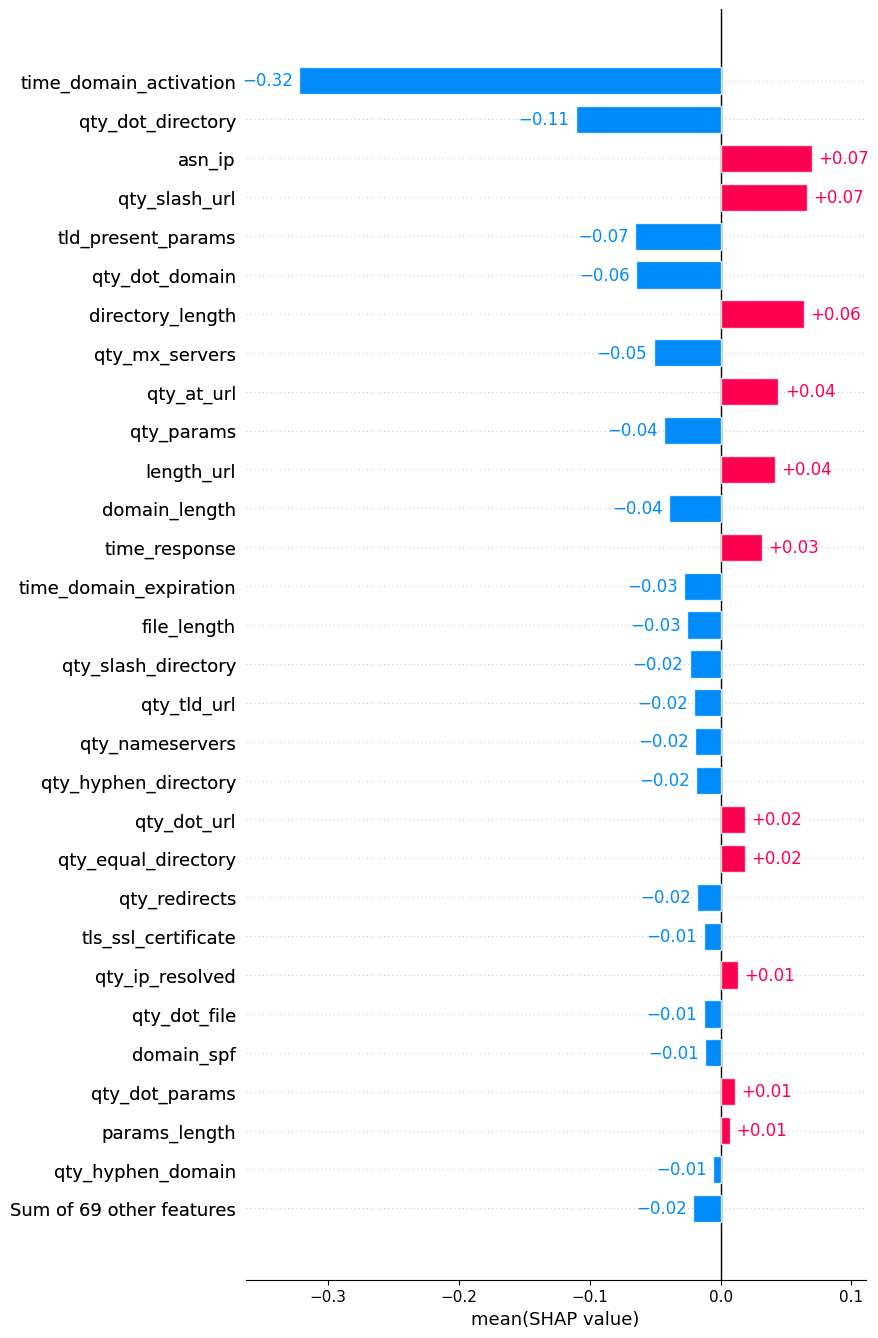}
    %\vspace{-3.5mm}
    \caption{SHAP bar plot of Exp-1}
    %\vspace{-1mm}
    \label{fig:exp1}
\end{figure}

%\vspace{-0.75mm}
\subsubsection{Exp-1 Feature Explanation}
Fig. \ref{fig:exp1} shows the top 30 influential features where we observe that 19 features are contributing negatively (blue bars), meaning on average, these features have contributed towards the {\em benign} class. Some of the highlighted negative contributing features are {\tt time\_domain\_activation} ($-0.32$), {\tt qty\_dot\_directory} ($-0.11$), {\tt tld\_present \_params} ($-0.07$), {\tt qty \_dot\_domain} ($-0.06$), {\tt qty\_mx\_servers} ($-0.05$), etc. From these 19 features, 3 are from the common feature list $F_c$ (\textit{$f_3$}, \textit{$f_7$} and \textit{$f_9$}). Figure \ref{fig:exp1} also shows that from the remaining 11 positively contributing features, top 4 are {\tt qty\_slash\_url} (\textit{$f_6$}) (+0.07),{\tt qty\_at\_url} (\textit{$f_{18}$}) (+0.04), {\tt length\_url} (\textit{$f_{13}$}) (+0.04), and {\tt qty\_dot\_url} (\textit{$f_1$}) (+0.02). % $f_1$, ,  and $f_{18}$). Some of the positively contributing features are %\textit{asn\_ip} (0.07), , \textit{directory\_length} (0.06),  etc.

% \begin{figure}[!ht]
%     \centering
%     \includegraphics[width=0.85\columnwidth]{Figure/exp1_summary.png}
%     \captionof{figure}{SHAP summary plot of Exp-1}
%     \label{fig:exp1_sum}
% \end{figure}

% \begin{figure}[!ht]
%     \centering
%     \includegraphics[width=0.85\columnwidth]{Figure/exp2_summary.png}
%     \caption{SHAP summary plot of Exp-2}
%     \label{fig:exp2_sum}
% \end{figure}

% \begin{figure*}[!htbp]
% \centering
% \subcaptionbox{\label{global}}{\includegraphics[width=0.5\textwidth]{Figure/exp1.png}}%
% \hfill % <-- Seperation
% \subcaptionbox{\label{local_1}}{\includegraphics[width=0.5\textwidth]{Figure/exp2.png}}%
% \caption{SHAP bar Plot of exp-1 and exp-2}
% \label{explanation}
% \end{figure*}

\begin{figure}[!t]
    \centering
    \includegraphics[width=1\columnwidth, height=1.25\linewidth]{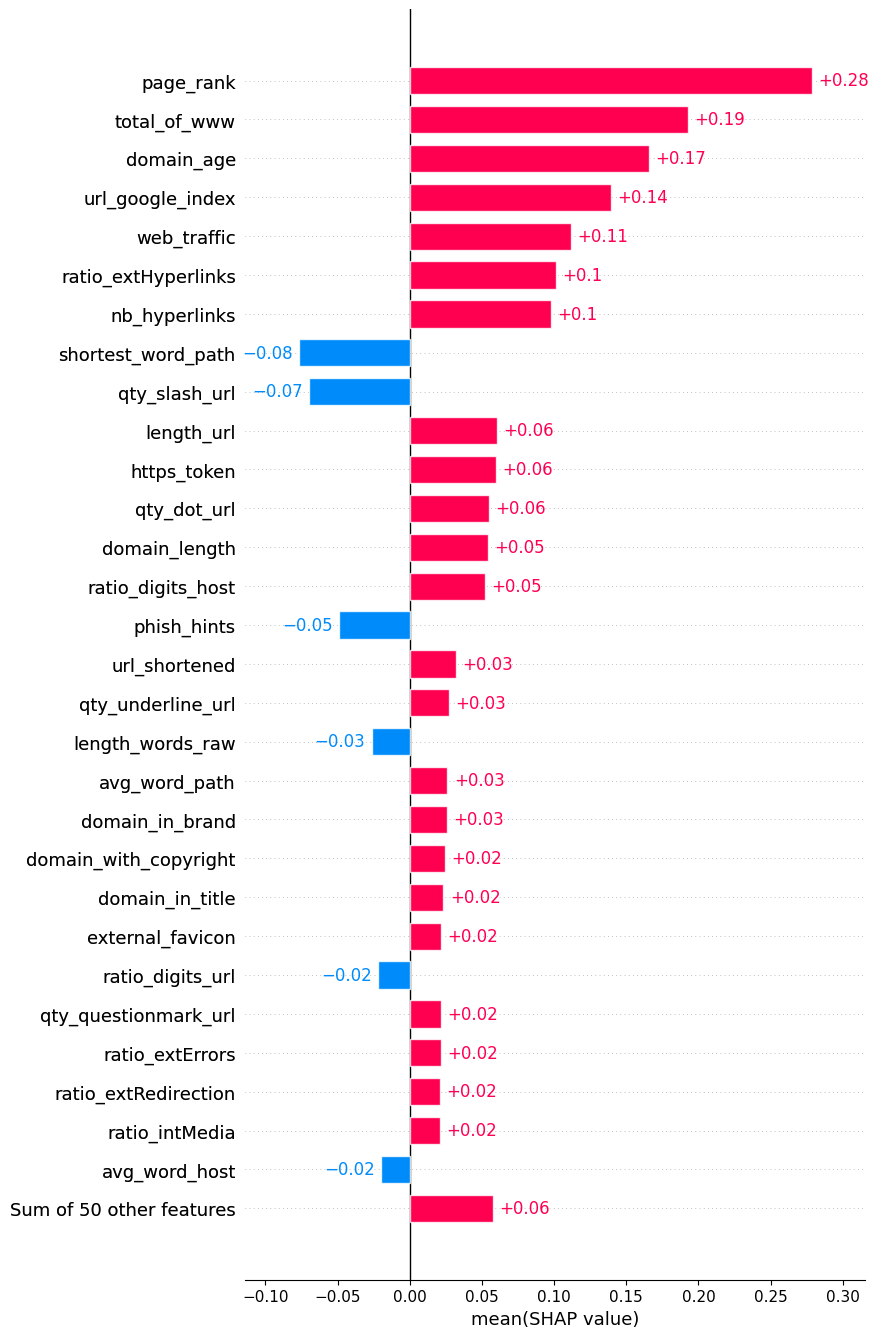}
    \caption{SHAP bar plot of Exp-2}
    %\vspace{-1mm}
    \label{fig:exp2}
\end{figure}

\begin{figure*}[!t]
\centering
\subcaptionbox{\label{global}}{\includegraphics[width=0.485\textwidth, height=0.40\linewidth]{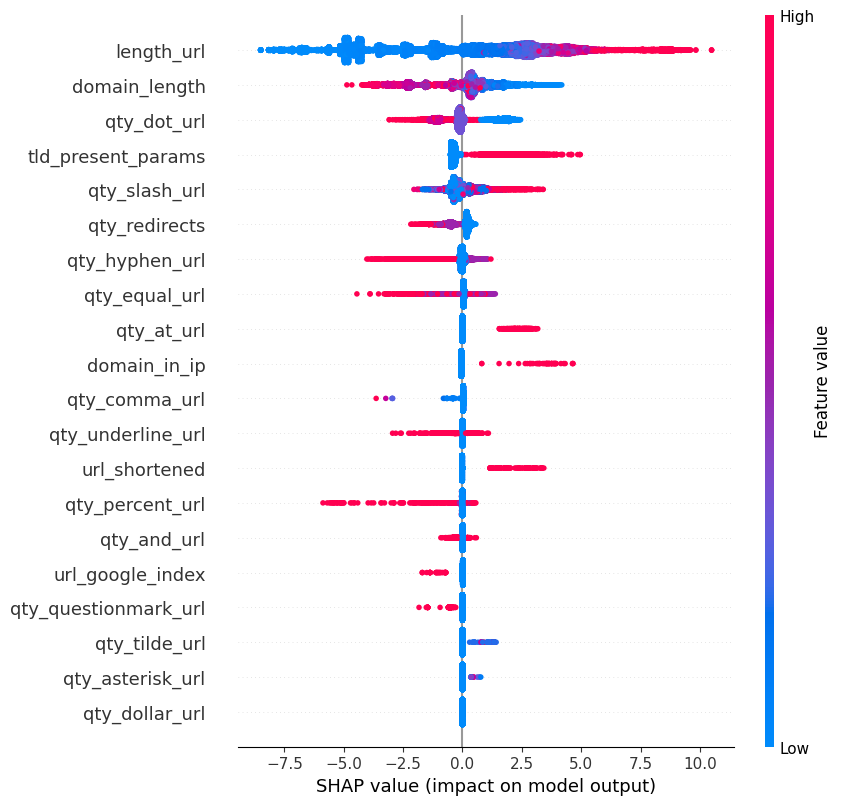}}%
\hfill % <-- Seperation
\subcaptionbox{\label{local_1}}{\includegraphics[width=0.485\textwidth, height=0.40\linewidth]{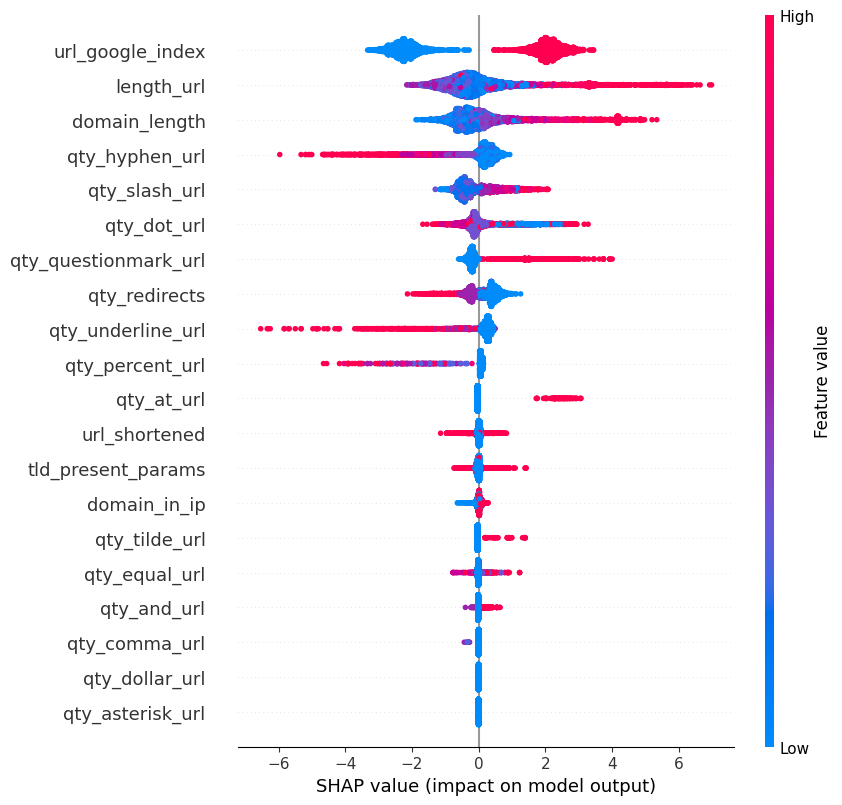}}%
\caption{SHAP summary plot for Exp-3 [6(a) on the left] and Exp-4 [6(b) on the right]}
\label{fig:sum_explanation}
\end{figure*}

%\vspace{-1mm}
\subsubsection{Exp-2 Feature Explanation}

Fig. \ref{fig:exp2} shows that $24$ out of the top $30$ most influential features are positively contributing (red bars). Some of the notable features are {\tt page\_rank} ($+0.28$), {\tt total\_of\_www} ($+0.19$), {\tt domain\_age} ($+0.17$), {\tt url\_google\_index} ($+0.14$) etc. From these $24$ positively contributing features, $7$ are in the $F_c$. \ignore{such as {\tt qty\_dot\_url} (\textit{$f_1$}), {\tt domain\_length} (\textit{$f_3$}), {\tt url\_google\_index}  (\textit{$f_4$}),{\tt url\_shortened} (\textit{$f_8$}), {\tt qty\_underline\_url} (\textit{$f_{12}$}), {\tt length\_url} (\textit{$f_{13}$}) and {\tt qty\_questionmark\_url} (\textit{$f_{16}$})} Interestingly, we can clearly identify that a particularly common feature {\tt domain\_length} (\textit{$f_3$}) is contributing positively in this case, which has shown a negative contribution in Exp-1 scenario. %but in this case we find that it is acting in opposite direction. 
From the remaining $6$ negatively contributing features, some notable ones are {\tt shortest\_word\_path} ($-0.08$), {\tt qty\_slash\_url} ($-0.07$), {\tt phish\_hints} ($-0.05$), {\tt length\_words\_raw} etc. Here, the contribution direction of feature {\tt qty\_slash\_url} (\textit{$f_6$}), which is from the common features, is also completely opposite of the observation in Exp-1 scenario. Another interesting insight is that the relative order of the common features based on the most influential contribution is also altered in Exp-2 compared to Exp-1. For example, in Exp-1 feature {\tt url\_google\_index} (\textit{$f_4$}) is not even in the top 30. However, in Exp-2 it is ranked as the $4$-th most contributing feature. From Exp-1 and Exp-2 observations, we have the following insights.

\textbf{Insight 1:} From the top 30 most impactful features, the majority are unique features, meaning they are only present in one dataset, which shows the dataset-dependent features in phishing detection. % which is a solid indicator of dataset dependency.

\textbf{Insight 2:}
    Even if common features are present in multiple phishing datasets, these features can contribute for predicting phishing class in one dataset while contribute for predicting benign class in the other dataset context.
 % which is a solid indicator of dataset dependency.

\textbf{Insight 3:}
    A particular feature may contribute positively for phishing detection in one dataset while the same feature may impact negatively in another dataset scenario.

%\vspace{-1mm}
\subsubsection{Exp-3 Feature Explanation }

% \begin{figure}[!b]
%     \centering
%     \includegraphics[width=1\columnwidth]{Figure/exp3.png}
%     \caption{SHAP bar plot of Exp-3}
%     \label{fig:exp3}
% \end{figure}

In this experiment scenario, we are using the 20 common features where training and testing are both conducted on dataset $D_1$. For this and the next experiment, we choose to depict the summary plot as it will show us the deviation in the value ranges for both dtasets. \ignore{Here, if we see the ranking of the features from figure \ref{fig:exp3}, the top positively contributing ones are {\tt qty\_at\_url} ($+0.07$), {\tt qty\_slash\_url} ($+0.06$), {\tt domain\_length} ($+0.05$), {\tt lenght\_url} ($+0.04$), {\tt qty\_comma\_url} ($+0.03$). 
\ignore{
and {\tt url\_google\_index} ($+0.01$)} On the other hand, the notable negatively contributing features are {\tt tld\_present \_params} ($-0.12$), {\tt qty\_redirects} ($-0.04$), {\tt domain\_in\_ip} ($-0.03$), {\tt qty \_percent\_url} ($-0.02$) and {\tt qty\_underline\_url} ($-0.01$). In comparison with Exp-1, all the features' impact directions are the same except for the {\tt domain\_length} (\textit{$f_3$}), and the SHAP values are very close for the other common features.}
From the summary plot Fig. \ref{fig:sum_explanation}(a), we can see that higher value of features {\em qty\_percent\_url}, {\em tld\_present\_params}, {\em qty\_slash\_url}, {\em qty\_at\_url}, {\em domain\_in\_ip}, \\{\em url\_shortened} and in addition, lower value of features {\em domain\_length}, {\em qty\_dot\_url}, {\em qty\_redirects}, {\em qty\_hyphen\_url}, {\em url\_google\_index}, \\{\em qty\_questionmark\_url} are contributing more on phishing prediction.%\footnote{add the full feature names here in the same text formatting as others... {\color{blue}done}}

%\vspace{-1mm}
\subsubsection{Exp-4 Feature Explanation }

% \begin{figure}[!t]
%     \centering
%     \includegraphics[width=1\columnwidth]{Figure/exp4.png}
%     \caption{SHAP bar plot of Exp-4}
%     \label{fig:exp4}
% \end{figure}

In this scenario, we are using the same common 20 features as Exp-3 but on a different dataset $D_2$. We expect that the apparent ranking order of features' contribution will be similar as both datasets share these same features. However, we observe the opposite of it.
The summary plot in Fig. \ref{fig:sum_explanation}(b) shows that higher values of features {\tt url\_google\_index}, {\tt qty\_percent\_url}, {\tt domain\_length}, {\tt qty\_slash\_url}, {\tt qty\_questionmark\_url}, {\tt qty\_at \_url} and lower values of features {\tt qty\_dot\_url}, {\tt qty\_redirects}, {\tt qty\_hyphen\_url},  {\tt qty\_underline\_url} are contributing more on phishing prediction. %\footnote{put the feature names instead of numbers for more clarification to the reader/reviewer! {\color{blue} done}}. 
This \textbf{answers RQ1} and unlike in Exp-3, we see that the lower value of features {\tt domain\_length}, {\tt qty\_questionmark \_url} contribute more to phishing prediction. From these two features, $f_3$ or {\tt domain\_length} is the most influential cause it's the highest ranked feature in Exp-4 and $4$-th ranked feature in Exp-3. Thus, the deviations between both Exp-3 and Exp-4 scenarios even when the same features are used for training and testing, raised a valid trust concern for features across different datasets. We can draw the following insights from both Exp-3 and Exp-4. 

\textbf{Insight 4:}
    The relative orders of the features' contributions in various dataset-agnostic experiments can vary vastly.

\textbf{Insight 5:}
    The value range of a particular feature may impact the prediction completely differently in multiple datasets.

\begin{figure}[!t]
    \centering
    \includegraphics[width=1\columnwidth, height=1\linewidth]{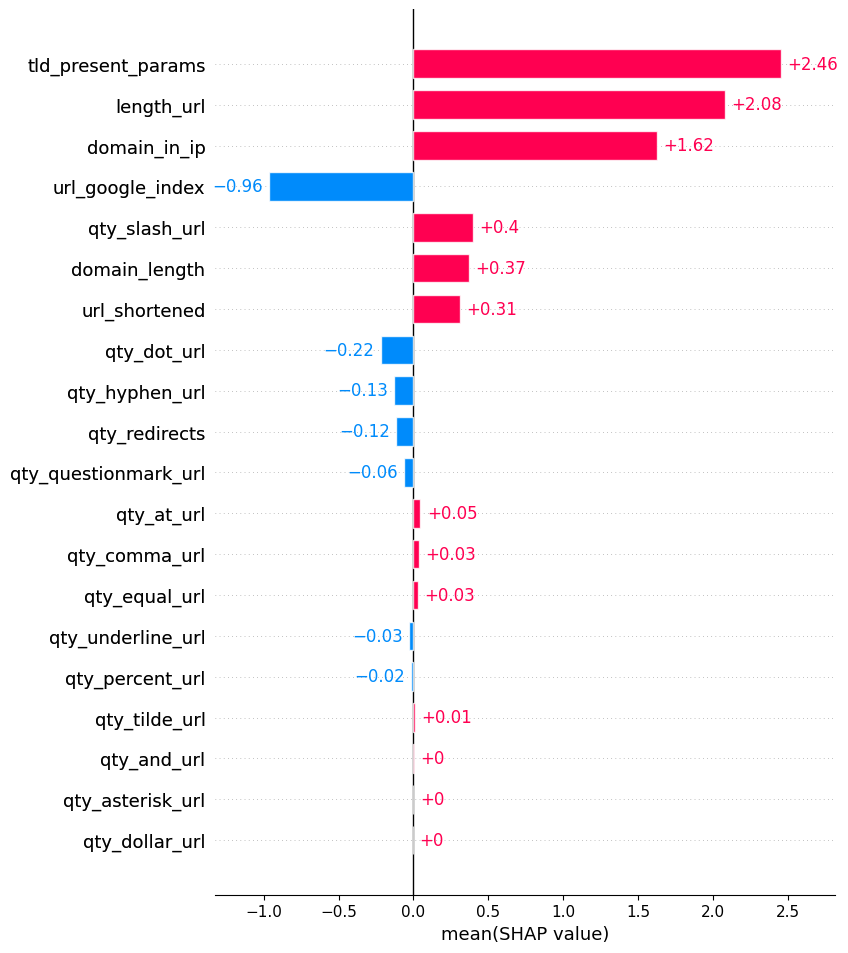}
    \caption{SHAP bar plot of Exp-5}
    %\vspace{-1mm}
    \label{fig:exp5}
\end{figure}

%\vspace{-1mm}
\subsubsection{Exp-5 Feature Explanation}
In this experiment scenario, we use the model that is trained on dataset $D_1$, but test it on the test portion of dataset $D_2$ as they both contain these 20 common features. This experiment help us \textbf{answering RQ2} in particular as we would learn how well or badly the model performs when the same dataset is not used in both the training and testing portions. %some additional data can be discovered if a model trained on one phishing website dataset effectively predict a different phishing website dataset. 
We observe from Fig. \ref{fig:exp5} that the impactful features with positive feature contributions are {\tt tld\_present\_params} ($+2.46$), {\tt length\_url} ($+2.08$), {\tt domain\_in\_ip} ($+1.62$), {\tt qty\_slash\_url} ($+0.4$), {\tt domain\_length} ($+0.37$) and {\tt url\_shortened} ($+0.31$). 
\ignore{In both Exp-3 and Exp-4 (when the same dataset is used for both training and testing), the contribution of feature {\tt tld\_present\_params} is in the negative direction. However, in this scenario it is surprisingly changed in the positive direction and in particular has the highest feature contribution among all the 20 features. A similar thing is observed for feature {\tt domain\_in\_ip} as it has the $3$-rd highest positive feature contributions in this scenario while in Exp-3 and Exp-4 cases this feature has negative and zero SHAP values, respectively.} 
Among the negatively contributing features, {\tt url\_google\_index} ($-0.96$), \\{\tt qty\_dot\_url} ($-0.22$), {\tt qty\_hyphen\_url} ($-0.13$), {\tt qty\_redirects} ($-0.12$) and {\tt qty\_questionmark\_url} ($-0.06$) are the highlighting ones. Interestingly, the direction of contributions for the features {\tt url\_google\_index} and {\tt qty\_dot\_url} are changed compared to Exp-4. \ignore{even though in both of these scenarios the test dataset is exactly the same and the SHAP is generated on the test samples. Lastly, feature {\tt qty\_redirects} being the third most positively contributing feature in Exp-4, is ranked at a much lower $10$-th position here.}

%\vspace{-0.75mm}
\subsubsection{Exp-6 Feature Explanation }
\begin{figure}[!t]
    \centering
    \includegraphics[width=1\columnwidth, height=1\linewidth]{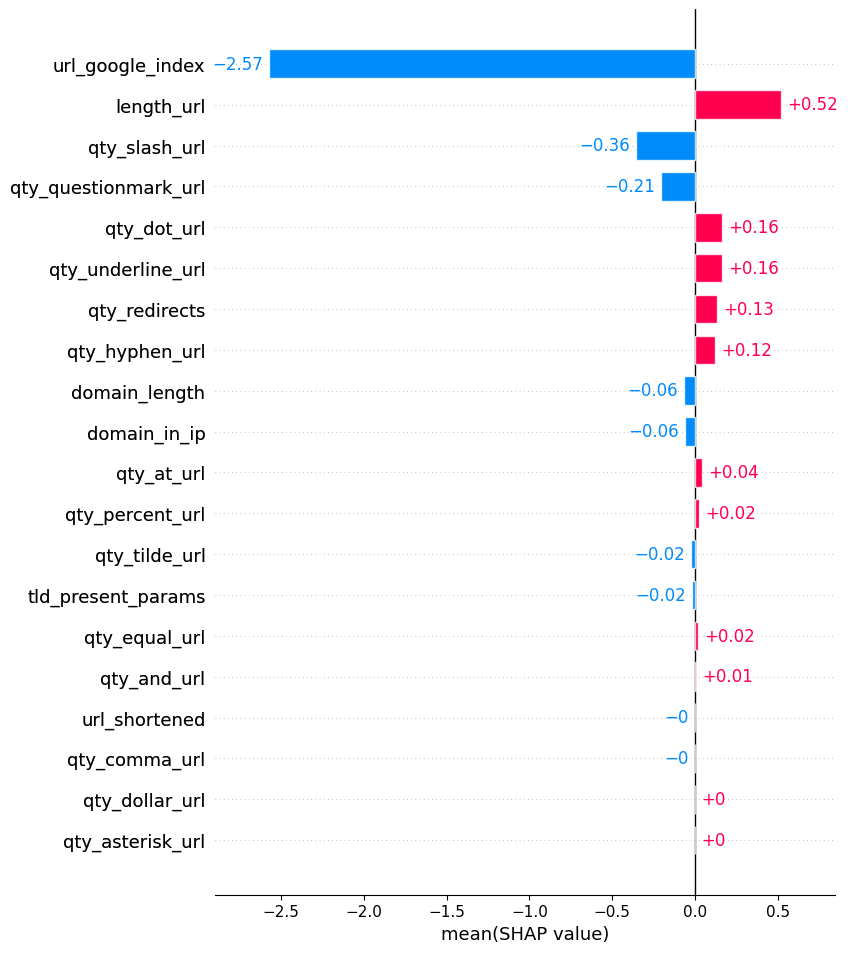}
    %\vspace{-1mm}
    \caption{SHAP bar plot of Exp-6}
    %\vspace{-1mm}
    \label{fig:exp6}
\end{figure}

In this experiment setup, we do the reverse scenario of Exp-5 where the model that is trained on dataset $D_2$, but we test it on the test portion of dataset $D_1$. From Fig. \ref{fig:exp6}, we observe that {\tt url\_google\_index} ($-2.57$) is the most influential feature with negative contributions though the same feature has positive contributions for single dataset train-test scenarios in both Exp-3 and Exp-4. In this experiment the other negatively contributing features are {\tt qty\_slash\_url} ($-0.36$), {\tt qty\_questionmark\_url} ($-0.21$), {\tt domain\_length} ($-0.06$) and {\tt domain\_in\_ip} ($-0.06$). On the contrary, some top positively contributing features are {\tt length\_url} ($+0.52$), {\tt qty\_dot\_url} ($+0.16$), {\tt qty\_underline\_url} ($+0.16$), \\{\tt qty\_redirects} ($+0.13$) and {\tt qty\_hyphen\_url} ($+0.12$). %\footnote{must address: what are the insights from this Exp-5 and Exp-6 scenarios? I the current description of Exp-6 I only see the feature contributions of positive and negative directions...but what does that infer?}

\textbf{Insight 6:}
    When we train on one dataset and test on the other dataset, the model performance degrades significantly and feature contributions on the test set deviate remarkably from the scenario where training and testing has been conducted on the same dataset. %has been used for both training and testing. 
    %In different train-test environment, feature contribution is not coherent with the model's actual train-test setup (Feature importance got changed in real-world application compared with the lab environment setup).

\subsubsection{Exp-7 Feature Explanation}

\begin{figure}[!t]
    \centering
    \includegraphics[ width=1\columnwidth, height=1\linewidth]{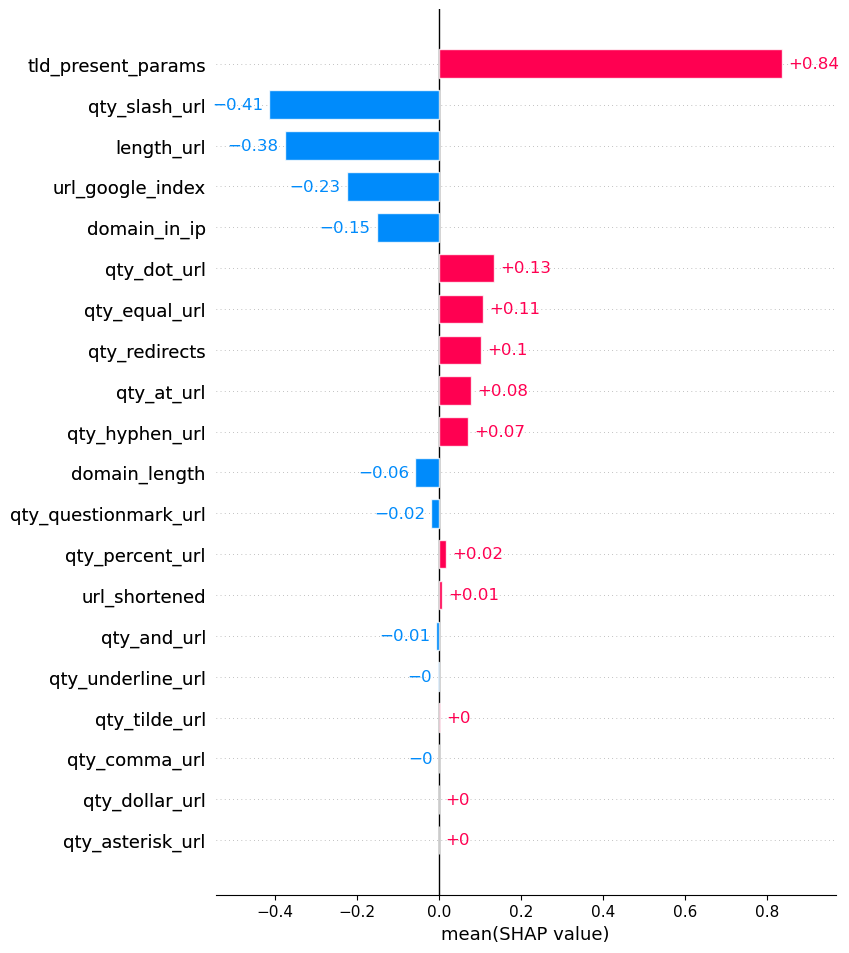}
    \caption{SHAP bar plot of Exp-7.1}
    \label{fig:exp7}
\end{figure}

%\vspace{-.5em}

These sub-experiments help us \textbf{answering RQ3}. Here, we use the mixed training and testing dataset $D_{merge}$. Here, we observe from Fig. \ref{fig:exp7} that for Exp-7.1, feature {\tt length\_url} ($-0.38$) is contributing negatively, meaning contributing more towards {\em benign} class, while in all previous experiment scenarios, this specific feature has contributed positively to detect phishing URLs. %meaning highly effective impacting one meaning it's helping more on legitimate prediction. 
Some negatively contributing features are {\tt url\_google\_index} ($-0.23$), {\tt qty\_slash\_url},($-0.41$), {\tt domain\_in\_ip} ($-0.15$) and {\tt domain\_length} ($-0.06$). On the other hand, significant positively contributing features are {\tt tld\_present\_params} ($+0.84$),  {\tt qty\_dot\_url} ($+0.13$), {\tt qty \_equal \_url} ($+0.11$) and {\tt qty\_redirects} ($+0.1$). \ignore{, {\tt qty\_at\_url} ($+0.08$) and {\tt qty\_hyphen\_url} ($+0.07$).} We can also see that the plot is quite similar to the Exp-6 plot in terms of features' contribution direction (positive or negative) except a very few features like {\tt length\_url} and {\tt tld\_present\_params}. The SHAP plot results from Exp-7.2 and Exp-7.3 are also coherent with Exp-7.1 but not presented for space constraints. %\textbf{These insights help us answer RQ3}.%\footnote{must address: same at previous Exp-6, what are the insights from here? why we test with merge dataset? need to draw from insights to make use of this experiment}
%\footnote{may be added if time allows: can you also test on $D_1$ test data and $D_2$ test data separately; we can show if the mixed training improve individual testing scenario...you can call it 7.2 and 7.3 ..and the current one as 7.1}

% \subsubsection{Exp-7.2 and Exp-7.3 Feature Explanation }

% \begin{figure}[!t]
%     \centering
%     \includegraphics[width=1\columnwidth]{Figure/exp7.2.png}
%     \caption{SHAP bar plot of Exp-7.2}
%     \label{fig:exp7.2}
% \end{figure}

% In comparison with experiment 7.1, Exp-7.2 shows (in Fig. \ref{fig:exp7.2}) an almost identical feature contribution bar plot. While, the mean SHAP bar plot for Exp-7.3, shown in Fig. \ref{fig:exp7.3}, portrays an opposite directional plot for features' contributions, meaning the features that contributed positively in Exp-7.1 are impacting negatively in Exp-7.3 scenario. However, in all these experiments (Exp-7.1, 7.2, 7.3), the relative order of the features in terms of overall impact on the model's prediction has been consistent. %changed significantly and which is a good indicator that merging dataset can create a monolithic set of contributing features. 

% \begin{figure}[!t]
%     \centering
%     \includegraphics[width=1\columnwidth]{Figure/exp7.3.png}
%     \caption{SHAP bar plot of Exp-7.3}
%     \label{fig:exp7.3}
% \end{figure}

\textbf{Insight 7:}
    Merging datasets for training can improve the model performance significantly even for diverse testing dataset scenarios and shows more consistent behaviors from the features' contribution and rank orders. 

%\vspace{-2.5mm}
\section{DISCUSSION AND LIMITATION}
\label{sec:discuss}
%\vspace{-0.3em}
\ignore{The experimental results provide insight into the global importance of various features in the detection of phishing URLs among different individual and mixed dataset scenarios.} %The SHAP global bar plot visualizes the impact of each feature on the model's output. 
Amongst the 20 common features across dataset $D_1$ and $D_2$, these following six features- {\tt qty\_dot\_directory}, {\tt qty \_slash \_url}, {\tt url\_google\_index}, {\tt length\_url}, {\tt tld\_present\_params}, and \\{\tt domain\_length} play a major role in the model's predictions in all of the experiment scenarios. However, the contribution orders of these features changes depending on the dataset that has been used for the training and testing. \ignore{We also see that some features impacted positively in several experiments while the same features contributed negatively in other experiment scenarios with a different individual or mixed dataset.} Additionally, we observe the deviations where in a particular experiment a specific feature's higher range value is impacting more on phishing detection while the lower range value of that same feature contributes more in another experiment setup. This deviation can cause significant detection errors in practice when a new type of data is introduced for prediction in a pre-trained phishing detection model. \ignore{Moreover, the adversaries can easily avoid detection from this kind of model when they know which features they need to manipulate to change the model decision from phishing to benign class.} % maybe a result of reverse attack of the adversaries that they may gain previous knowledge for that particular features and change their value to avoid detection or the datasets are not generalized. 
Thus, we recommend developing a more complete publicly available bench-mark dataset from the research community for comparing model's performances as well as a comprehensive feature list. We also recommend keeping the dataset live and updated as data trends may change from time to time as attackers adjust their techniques and tactics through adversarial attacks.  

Our experiments find that the model that works well when trained and tested on the same dataset, may not be effective when trained on one dataset and tested on another. This results into a very low accuracy for Exp-5 and Exp-6 scenarios, which infers that the model is not generalized well for phishing detection in practice and may only be effective in that particular dataset scenario. Initially, we have assumed that merging dataset $D_1$ and $D_2$ (in Exp-7) would result in better performance than Exp-5 or Exp-6 scenarios. Interestingly, that is turned out to be true as the trained model on the merged dataset capable of performing well on the test portion of not only the merged dataset but also the test portion of individual datasets $D_1$ and $D_2$. Thus, we recommend training on diverse datasets in practice for designing effective phishing URL or website detection models rather than using any singular dataset. %Furthermore, the results reveal that combining the two datasets even worsen model performance, as evidenced by lower accuracy despite having equal number of random instances from $D_1$ and $D_2$ present in the $D_{merge}$.
%\vspace{-3mm}
\subsection{Limitations}
%\vspace{-0.3em}
Even though we shed light on the phishing detection features' deviation of effectiveness across datasets, there remain some limitations in the present study. %One limitation of this study is that it only considers global feature relevance. While this provides a comprehensive perspective of feature relevance, 
\textbf{First}, the study does not take into consideration each feature's instance-specific contributions (local explanation) for identifying which features are deviating in the case of an individual test instance when trained on one dataset and tested on another. %Future research could investigate the impact of local features' impact investigated on several individual instances in order to acquire a more detailed understanding of the model's predictions. 
\textbf{Second}, we rely on the features' SHAP values when interpreting the model. While SHAP gives a single measure of feature relevance, it makes the assumption that features are independent, which may not be true always. %\textbf{Third}, we only used two datasets in this study and maybe further datasets can be incorporated to generalize the concepts. %It could be interesting if the merged dataset training can also be validated to provide reasonably good results for another third dataset containing the same common features and characteristics for testing. 
\textbf{Third}, We also have not considered any datasets with shared visual similarity features because of scarcity of such features. 
\textbf{Fourth}, the common phishing versus benign URL classification features that we have analyzed are limited to mostly lexical features as they are more commonly found across multiple datasets. \ignore{However, we understand that web page content-based features can be effective in phishing detection and if multiple differently sourced datasets have these content-based features in common, we can analyze those as well to understand if they show a similar nature to the lexical ones in this paper.} \textbf{Fifth}, we have not considered the deep learning methods which can be considered in future studies. %give more relevant results.    %the meaning of SHAP values is relative and is determined by the model's output scale and distribution, as well as the SHAP values for other characteristics.

\ignore{
{\color{red}
Reviewers comments/concerns:

1. I would have liked to know how the dataset-dependent features map to the different datasets more clearly, to perhaps better understand why Exp-2 / Exp-4 outperformed Exp 7.1-3
{\color{blue} Recommendation: Highlight the difference in explanation or discussion}

2. Since the common features listed in Table II have different scales (e.g., some are boolean or percentage), examining the mean values in the same figure (Fig. 2 and FIg. 3) stretched to the largest scale may be ill-advised, as it makes it difficult to discern differences between the datasets for small-valued features. Additionally, both Figure 4 and Figure 5 do not appear to be referenced in the main text.
{\color{blue} Recommendation: Separate figures for boolean feature versus natural number valued features; Fig 4 and Fig 5 reference should be checked.}

3. In Section IV-C (Train and Test with Supervised XAI Models), it looks like only more "traditional" ML models (XGB, LR, DR, RF fNB, etc) are selected. It would be awesome if the authors can discuss some more "modern" models like CNN, RNN, LSTM, and even LLM and RAG. {\color{blue} Recommendation: Accept it as a limitation and mention as future work}

4. {\color{blue} Recommendation: Make the paper fit within 8 pages including references. Reduce related work section; reduce Sec 3.1; Remove fig 4; Review Exp to remove if something is redundant or less interesting!}
}
}

%\vspace{-2mm}
\section{CONCLUSION}
\label{sec:conclusion}
%\vspace{-.2em}
\ignore{This paper has made substantial progress in comprehending the global significance of numerous features and their connection in the detection of phishing URLs from the two different datasets using XAI.}

The usage of Explainable AI has allowed us to establish variables which are most important in detecting phishing websites, as well as how these features differ between datasets. \ignore{The study also offered light on the problems encountered while assessing the model between various datasets. It has highlighted the importance of strong, dataset-agnostic phishing detection models, as well as the potential of explainable AI as a validation tool to achieve this goal.} %Using SHAP, a strong method for understanding model predictions, the impact of each feature on the model's output across two different datasets is able to be assessed. 
The analysis discovered critical variables that contribute significantly to the model's predictions in several experiment scenarios. While the model performs well when trained and evaluated on the same dataset, its effectiveness is reduced when the training and test dataset are different even though they share the same features. This experimental evidences stress the need of training on different and representative datasets to improve model generalizability in practice before claiming a phishing detection model as effective with certain features. Thus, we can evidently state that phishing URL detection model that trained on singular dataset specifically using the lexical features can not be trusted across diverse datasets even if the features have very close data distributions. We recommend using multiple diverse merged dataset as a better method to use as training for the phishing detection model and always use explainable methods for AI/ML model's verification and trustworthiness for further decision-making on the outcome. %In future, more datasets can be incorporated with more diverse feature lists.
The detail implementation code and dataset are shared in the following Github repository (\url{https://anonymous.4open.science/r/Deviation-in-Feature-Contribution-7760/}) for reproducing the experiments. %{\color{red}The study also found that merging the datasets can improve the model's performance.}\footnote{not true at this point} %However, the combined model's performance differs depending on the test dataset, emphasizing the need for additional research into this area.

%The findings have offered a full understanding of the common and distinctive features found in numerous datasets utilized for phishing detection, as well as their contribution to the detection process. 
%Despite these major contributions, the study has indicated possible topics for further investigation. The study emphasized the need for greater research into the impact of local features in order to have a more detailed grasp of the model's predictions. Furthermore, while using SHAP values to understand the model is useful, it implies feature independence and relative interpretation, which may not always be the case.

%assistive technologies to
%\begin{acks}
%ack goes here...
%\end{acks}

\bibliographystyle{ACM-Reference-Format}
\bibliography{sample-base}


\begin{thebibliography}{41}

%%% the Web address.

\ifx \showCODEN    \undefined \def \showCODEN     #1{\unskip}     \fi
\ifx \showDOI      \undefined \def \showDOI       #1{#1}\fi
\ifx \showISBNx    \undefined \def \showISBNx     #1{\unskip}     \fi
\ifx \showISBNxiii \undefined \def \showISBNxiii  #1{\unskip}     \fi
\ifx \showISSN     \undefined \def \showISSN      #1{\unskip}     \fi
\ifx \showLCCN     \undefined \def \showLCCN      #1{\unskip}     \fi
\ifx \shownote     \undefined \def \shownote      #1{#1}          \fi
\ifx \showarticletitle \undefined \def \showarticletitle #1{#1}   \fi
\ifx \showURL      \undefined \def \showURL       {\relax}        \fi
% invisible to TeX
\providecommand\bibfield[2]{#2}
\providecommand\bibinfo[2]{#2}
\providecommand\natexlab[1]{#1}
\providecommand\showeprint[2][]{arXiv:#2}

\bibitem[Abdelhamid et~al\mbox{.}(2017)]%
        {abdelhamid2017phishing}
\bibfield{author}{\bibinfo{person}{Neda Abdelhamid}, \bibinfo{person}{Fadi Thabtah}, {and} \bibinfo{person}{Hussein Abdel-Jaber}.} \bibinfo{year}{2017}\natexlab{}.
\newblock \showarticletitle{Phishing detection: A recent intelligent machine learning comparison based on models content and features}. In \bibinfo{booktitle}{\emph{2017 IEEE international conference on intelligence and security informatics (ISI)}}. IEEE, \bibinfo{pages}{72--77}.
\newblock


\bibitem[Aljofey et~al\mbox{.}(2020)]%
        {aljofey2020_cnn_based_phishing}
\bibfield{author}{\bibinfo{person}{Ali Aljofey}, \bibinfo{person}{Qingshan Jiang}, \bibinfo{person}{Qiang Qu}, \bibinfo{person}{Mingqing Huang}, {and} \bibinfo{person}{Jean-Pierre Niyigena}.} \bibinfo{year}{2020}\natexlab{}.
\newblock \showarticletitle{An effective phishing detection model based on character level convolutional neural network from URL}.
\newblock \bibinfo{journal}{\emph{Electronics}} \bibinfo{volume}{9}, \bibinfo{number}{9} (\bibinfo{year}{2020}), \bibinfo{pages}{1514}.
\newblock


\bibitem[Aljofey et~al\mbox{.}(2022)]%
        {aljofey2022effective_html_url}
\bibfield{author}{\bibinfo{person}{Ali Aljofey}, \bibinfo{person}{Qingshan Jiang}, \bibinfo{person}{Abdur Rasool}, \bibinfo{person}{Hui Chen}, \bibinfo{person}{Wenyin Liu}, \bibinfo{person}{Qiang Qu}, {and} \bibinfo{person}{Yang Wang}.} \bibinfo{year}{2022}\natexlab{}.
\newblock \showarticletitle{An effective detection approach for phishing websites using URL and HTML features}.
\newblock \bibinfo{journal}{\emph{Scientific Reports}} \bibinfo{volume}{12}, \bibinfo{number}{1} (\bibinfo{year}{2022}), \bibinfo{pages}{8842}.
\newblock


\bibitem[Bouijij and Berqia(2021)]%
        {b4}
\bibfield{author}{\bibinfo{person}{Habiba Bouijij} {and} \bibinfo{person}{Amine Berqia}.} \bibinfo{year}{2021}\natexlab{}.
\newblock \showarticletitle{Machine Learning Algorithms Evaluation for Phishing URLs Classification}.
\newblock \bibinfo{journal}{\emph{2021 4th International Symposium on Advanced Electrical and Communication Technologies (ISAECT)}} (\bibinfo{year}{2021}), \bibinfo{pages}{01--05}.
\newblock
\urldef\tempurl%
\url{https://api.semanticscholar.org/CorpusID:245881138}
\showURL{%
\tempurl}


\bibitem[Chawla et~al\mbox{.}(2002)]%
        {b5}
\bibfield{author}{\bibinfo{person}{N.~V. Chawla}, \bibinfo{person}{K.~W. Bowyer}, \bibinfo{person}{L.~O. Hall}, {and} \bibinfo{person}{W.~P. Kegelmeyer}.} \bibinfo{year}{2002}\natexlab{}.
\newblock \showarticletitle{SMOTE: Synthetic Minority Over-sampling Technique}.
\newblock \bibinfo{journal}{\emph{Journal of Artificial Intelligence Research}}  \bibinfo{volume}{16} (\bibinfo{date}{June} \bibinfo{year}{2002}), \bibinfo{pages}{321–357}.
\newblock
\showISSN{1076-9757}
\urldef\tempurl%
\url{https://doi.org/10.1613/jair.953}
\showDOI{\tempurl}


\bibitem[Chiew et~al\mbox{.}(2018)]%
        {chiew2018building}
\bibfield{author}{\bibinfo{person}{Kang~Leng Chiew}, \bibinfo{person}{Ee~Hung Chang}, \bibinfo{person}{C~Lin Tan}, \bibinfo{person}{Johari Abdullah}, {and} \bibinfo{person}{Kelvin Sheng~Chek Yong}.} \bibinfo{year}{2018}\natexlab{}.
\newblock \showarticletitle{Building standard offline anti-phishing dataset for benchmarking}.
\newblock \bibinfo{journal}{\emph{International Journal of Engineering \& Technology}} \bibinfo{volume}{7}, \bibinfo{number}{4.31} (\bibinfo{year}{2018}), \bibinfo{pages}{7--14}.
\newblock


\bibitem[Chinnasamy et~al\mbox{.}(2022)]%
        {b3}
\bibfield{author}{\bibinfo{person}{P. Chinnasamy}, \bibinfo{person}{N. Kumaresan}, \bibinfo{person}{R. Selvaraj}, \bibinfo{person}{S. Dhanasekaran}, \bibinfo{person}{K Ramprathap}, {and} \bibinfo{person}{Sruthi Boddu}.} \bibinfo{year}{2022}\natexlab{}.
\newblock \showarticletitle{An Efficient Phishing Attack Detection using Machine Learning Algorithms}. In \bibinfo{booktitle}{\emph{2022 International Conference on Advancements in Smart, Secure and Intelligent Computing (ASSIC)}}. \bibinfo{pages}{1--6}.
\newblock
\urldef\tempurl%
\url{https://doi.org/10.1109/ASSIC55218.2022.10088399}
\showDOI{\tempurl}


\bibitem[Das~Guptta et~al\mbox{.}(2024)]%
        {das2024modeling_hybrid_features}
\bibfield{author}{\bibinfo{person}{Sumitra Das~Guptta}, \bibinfo{person}{Khandaker~Tayef Shahriar}, \bibinfo{person}{Hamed Alqahtani}, \bibinfo{person}{Dheyaaldin Alsalman}, {and} \bibinfo{person}{Iqbal~H Sarker}.} \bibinfo{year}{2024}\natexlab{}.
\newblock \showarticletitle{Modeling hybrid feature-based phishing websites detection using machine learning techniques}.
\newblock \bibinfo{journal}{\emph{Annals of Data Science}} \bibinfo{volume}{11}, \bibinfo{number}{1} (\bibinfo{year}{2024}), \bibinfo{pages}{217--242}.
\newblock


\bibitem[El~Aassal et~al\mbox{.}(2020)]%
        {el2020depth}
\bibfield{author}{\bibinfo{person}{Ayman El~Aassal}, \bibinfo{person}{Shahryar Baki}, \bibinfo{person}{Avisha Das}, {and} \bibinfo{person}{Rakesh~M Verma}.} \bibinfo{year}{2020}\natexlab{}.
\newblock \showarticletitle{An in-depth benchmarking and evaluation of phishing detection research for security needs}.
\newblock \bibinfo{journal}{\emph{Ieee Access}}  \bibinfo{volume}{8} (\bibinfo{year}{2020}), \bibinfo{pages}{22170--22192}.
\newblock


\bibitem[Elsadig et~al\mbox{.}(2022)]%
        {elsadig2022intelligent_bert}
\bibfield{author}{\bibinfo{person}{Muna Elsadig}, \bibinfo{person}{Ashraf~Osman Ibrahim}, \bibinfo{person}{Shakila Basheer}, \bibinfo{person}{Manal~Abdullah Alohali}, \bibinfo{person}{Sara Alshunaifi}, \bibinfo{person}{Haya Alqahtani}, \bibinfo{person}{Nihal Alharbi}, {and} \bibinfo{person}{Wamda Nagmeldin}.} \bibinfo{year}{2022}\natexlab{}.
\newblock \showarticletitle{Intelligent deep machine learning cyber phishing url detection based on bert features extraction}.
\newblock \bibinfo{journal}{\emph{Electronics}} \bibinfo{volume}{11}, \bibinfo{number}{22} (\bibinfo{year}{2022}), \bibinfo{pages}{3647}.
\newblock


\bibitem[Feng and Yue(2020)]%
        {feng2020visualizing}
\bibfield{author}{\bibinfo{person}{Tao Feng} {and} \bibinfo{person}{Chuan Yue}.} \bibinfo{year}{2020}\natexlab{}.
\newblock \showarticletitle{Visualizing and interpreting rnn models in url-based phishing detection}. In \bibinfo{booktitle}{\emph{Proceedings of the 25th ACM Symposium on Access Control Models and Technologies}}. \bibinfo{pages}{13--24}.
\newblock


\bibitem[Group({[n.\,d.]})]%
        {phishtank_portal}
\bibfield{author}{\bibinfo{person}{Cisco Talos~Intelligence Group}.} \bibinfo{year}{[n.\,d.]}\natexlab{}.
\newblock \bibinfo{title}{PhishTank}.
\newblock \bibinfo{howpublished}{https://phishtank.org/}.
\newblock
\newblock
\shownote{(Accessed \ on \ 1 \ September, \ 2024)}.


\bibitem[Gupta et~al\mbox{.}(2021)]%
        {gupta2021novel}
\bibfield{author}{\bibinfo{person}{Brij~B Gupta}, \bibinfo{person}{Krishna Yadav}, \bibinfo{person}{Imran Razzak}, \bibinfo{person}{Konstantinos Psannis}, \bibinfo{person}{Arcangelo Castiglione}, {and} \bibinfo{person}{Xiaojun Chang}.} \bibinfo{year}{2021}\natexlab{}.
\newblock \showarticletitle{A novel approach for phishing URLs detection using lexical based machine learning in a real-time environment}.
\newblock \bibinfo{journal}{\emph{Computer Communications}}  \bibinfo{volume}{175} (\bibinfo{year}{2021}), \bibinfo{pages}{47--57}.
\newblock


\bibitem[Gupta et~al\mbox{.}(2023)]%
        {threatGPT_Maanak2023Access}
\bibfield{author}{\bibinfo{person}{Maanak Gupta}, \bibinfo{person}{Charankumar Akiri}, \bibinfo{person}{Kshitiz Aryal}, \bibinfo{person}{Eli Parker}, {and} \bibinfo{person}{Lopamudra Praharaj}.} \bibinfo{year}{2023}\natexlab{}.
\newblock \showarticletitle{From ChatGPT to ThreatGPT: Impact of Generative AI in Cybersecurity and Privacy}.
\newblock \bibinfo{journal}{\emph{IEEE Access}}  \bibinfo{volume}{11} (\bibinfo{year}{2023}), \bibinfo{pages}{80218--80245}.
\newblock
\urldef\tempurl%
\url{https://doi.org/10.1109/ACCESS.2023.3300381}
\showDOI{\tempurl}


\bibitem[Hannousse and Yahiouche(2021)]%
        {hannousse2021towards}
\bibfield{author}{\bibinfo{person}{Abdelhakim Hannousse} {and} \bibinfo{person}{Salima Yahiouche}.} \bibinfo{year}{2021}\natexlab{}.
\newblock \showarticletitle{Towards benchmark datasets for machine learning based website phishing detection: An experimental study}.
\newblock \bibinfo{journal}{\emph{Engineering Applications of Artificial Intelligence}}  \bibinfo{volume}{104} (\bibinfo{year}{2021}), \bibinfo{pages}{104347}.
\newblock


\bibitem[Jain and Gupta(2018)]%
        {jain2018phish}
\bibfield{author}{\bibinfo{person}{Ankit~Kumar Jain} {and} \bibinfo{person}{Brij~B Gupta}.} \bibinfo{year}{2018}\natexlab{}.
\newblock \showarticletitle{PHISH-SAFE: URL features-based phishing detection system using machine learning}. In \bibinfo{booktitle}{\emph{Cyber Security: Proceedings of CSI 2015}}. Springer, \bibinfo{pages}{467--474}.
\newblock


\bibitem[Karim et~al\mbox{.}(2023)]%
        {b2}
\bibfield{author}{\bibinfo{person}{Abdul Karim}, \bibinfo{person}{Mobeen Shahroz}, \bibinfo{person}{Khabib Mustofa}, \bibinfo{person}{Samir Brahim~Belhaouari}, {and} \bibinfo{person}{S~Ramana~Kumar Joga}.} \bibinfo{year}{2023}\natexlab{}.
\newblock \showarticletitle{Phishing Detection System Through Hybrid Machine Learning Based on URL}.
\newblock \bibinfo{journal}{\emph{IEEE Access}}  \bibinfo{volume}{PP} (\bibinfo{date}{01} \bibinfo{year}{2023}), \bibinfo{pages}{1--1}.
\newblock
\urldef\tempurl%
\url{https://doi.org/10.1109/ACCESS.2023.3252366}
\showDOI{\tempurl}


\bibitem[Lundberg and Lee(2017)]%
        {NIPS2017_7062_shap}
\bibfield{author}{\bibinfo{person}{Scott~M Lundberg} {and} \bibinfo{person}{Su-In Lee}.} \bibinfo{year}{2017}\natexlab{}.
\newblock \showarticletitle{A Unified Approach to Interpreting Model Predictions}.
\newblock In \bibinfo{booktitle}{\emph{Advances in Neural Information Processing Systems 30}}, \bibfield{editor}{\bibinfo{person}{I.~Guyon}, \bibinfo{person}{U.~V. Luxburg}, \bibinfo{person}{S.~Bengio}, \bibinfo{person}{H.~Wallach}, \bibinfo{person}{R.~Fergus}, \bibinfo{person}{S.~Vishwanathan}, {and} \bibinfo{person}{R.~Garnett}} (Eds.). \bibinfo{publisher}{Curran Associates, Inc.}, \bibinfo{pages}{4765--4774}.
\newblock
\urldef\tempurl%
\url{http://papers.nips.cc/paper/7062-a-unified-approach-to-interpreting-model-predictions.pdf}
\showURL{%
\tempurl}


\bibitem[Main(2023)]%
        {phishing_stat_M.K}
\bibfield{author}{\bibinfo{person}{Kelly Main}.} \bibinfo{year}{2023}\natexlab{}.
\newblock \bibinfo{title}{Phishing statistics by state in 2024}.
\newblock \bibinfo{howpublished}{https://www.forbes.com/advisor/business/phishing-statistics/}.
\newblock
\newblock
\shownote{(Accessed \ on \ 15 \ May, \ 2024)}.


\bibitem[Maroofi et~al\mbox{.}(2020)]%
        {comar_euro_sp2020_maroofi}
\bibfield{author}{\bibinfo{person}{Sourena Maroofi}, \bibinfo{person}{Maciej Korczyński}, \bibinfo{person}{Cristian Hesselman}, \bibinfo{person}{Benoît Ampeau}, {and} \bibinfo{person}{Andrzej Duda}.} \bibinfo{year}{2020}\natexlab{}.
\newblock \showarticletitle{COMAR: Classification of Compromised versus Maliciously Registered Domains}. In \bibinfo{booktitle}{\emph{2020 IEEE European Symposium on Security and Privacy (EuroS\&P)}}. \bibinfo{pages}{607--623}.
\newblock
\urldef\tempurl%
\url{https://doi.org/10.1109/EuroSP48549.2020.00045}
\showDOI{\tempurl}


\bibitem[Menon and Gressel(2021)]%
        {menon2021concept}
\bibfield{author}{\bibinfo{person}{Aditya~Gopal Menon} {and} \bibinfo{person}{Gilad Gressel}.} \bibinfo{year}{2021}\natexlab{}.
\newblock \showarticletitle{Concept drift detection in phishing using autoencoders}. In \bibinfo{booktitle}{\emph{Machine Learning and Metaheuristics Algorithms, and Applications: Second Symposium, SoMMA 2020, Chennai, India, October 14--17, 2020, Revised Selected Papers 2}}. Springer, \bibinfo{pages}{208--220}.
\newblock


\bibitem[Mourtaji et~al\mbox{.}(2021)]%
        {mourtaji2021hybrid_cnn}
\bibfield{author}{\bibinfo{person}{Youness Mourtaji}, \bibinfo{person}{Mohammed Bouhorma}, \bibinfo{person}{Daniyal Alghazzawi}, \bibinfo{person}{Ghadah Aldabbagh}, {and} \bibinfo{person}{Abdullah Alghamdi}.} \bibinfo{year}{2021}\natexlab{}.
\newblock \showarticletitle{Hybrid rule-based solution for phishing URL detection using convolutional neural network}.
\newblock \bibinfo{journal}{\emph{Wireless Communications and Mobile Computing}}  \bibinfo{volume}{2021} (\bibinfo{year}{2021}), \bibinfo{pages}{1--24}.
\newblock


\bibitem[Opara et~al\mbox{.}(2024)]%
        {opara2024look_url_html}
\bibfield{author}{\bibinfo{person}{Chidimma Opara}, \bibinfo{person}{Yingke Chen}, {and} \bibinfo{person}{Bo Wei}.} \bibinfo{year}{2024}\natexlab{}.
\newblock \showarticletitle{Look before You leap: Detecting phishing web pages by exploiting raw URL And HTML characteristics}.
\newblock \bibinfo{journal}{\emph{Expert Systems with Applications}}  \bibinfo{volume}{236} (\bibinfo{year}{2024}), \bibinfo{pages}{121183}.
\newblock


\bibitem[Preeti and Sharma(2023)]%
        {b8}
\bibfield{author}{\bibinfo{person}{Preeti} {and} \bibinfo{person}{Priti Sharma}.} \bibinfo{year}{2023}\natexlab{}.
\newblock \showarticletitle{A Detailed Analysis on Various Datasets using Machine learning and Deep Learning Techniques for Phishing URLs Detection}. In \bibinfo{booktitle}{\emph{2023 14th International Conference on Computing Communication and Networking Technologies (ICCCNT)}}. \bibinfo{pages}{1--10}.
\newblock
\urldef\tempurl%
\url{https://doi.org/10.1109/ICCCNT56998.2023.10307474}
\showDOI{\tempurl}


\bibitem[Pritom et~al\mbox{.}(2020)]%
        {pritom2020_covid_malwebsites}
\bibfield{author}{\bibinfo{person}{Mir Mehedi~Ahsan Pritom}, \bibinfo{person}{Kristin~M. Schweitzer}, \bibinfo{person}{Raymond~M. Bateman}, \bibinfo{person}{Min Xu}, {and} \bibinfo{person}{Shouhuai Xu}.} \bibinfo{year}{2020}\natexlab{}.
\newblock \showarticletitle{Data-Driven Characterization and Detection of COVID-19 Themed Malicious Websites}. In \bibinfo{booktitle}{\emph{2020 IEEE International Conference on Intelligence and Security Informatics (ISI)}}. \bibinfo{pages}{1--6}.
\newblock
\urldef\tempurl%
\url{https://doi.org/10.1109/ISI49825.2020.9280522}
\showDOI{\tempurl}


\bibitem[Pritom and Xu(2022)]%
        {pritom2022_cns_website_lawenforcement}
\bibfield{author}{\bibinfo{person}{Mir Mehedi~Ahsan Pritom} {and} \bibinfo{person}{Shouhuai Xu}.} \bibinfo{year}{2022}\natexlab{}.
\newblock \showarticletitle{Supporting Law-Enforcement to Cope with Blacklisted Websites: Framework and Case Study}. In \bibinfo{booktitle}{\emph{2022 IEEE Conference on Communications and Network Security (CNS)}}. \bibinfo{pages}{181--189}.
\newblock
\urldef\tempurl%
\url{https://doi.org/10.1109/CNS56114.2022.9947260}
\showDOI{\tempurl}


\bibitem[Rao et~al\mbox{.}(2020)]%
        {rao2020catchphish_url}
\bibfield{author}{\bibinfo{person}{Routhu~Srinivasa Rao}, \bibinfo{person}{Tatti Vaishnavi}, {and} \bibinfo{person}{Alwyn~Roshan Pais}.} \bibinfo{year}{2020}\natexlab{}.
\newblock \showarticletitle{CatchPhish: detection of phishing websites by inspecting URLs}.
\newblock \bibinfo{journal}{\emph{Journal of Ambient Intelligence and Humanized Computing}}  \bibinfo{volume}{11} (\bibinfo{year}{2020}), \bibinfo{pages}{813--825}.
\newblock


\bibitem[Roy et~al\mbox{.}(2024)]%
        {SP2024_Phishbot_chatbot}
\bibfield{author}{\bibinfo{person}{Sayak~Saha Roy}, \bibinfo{person}{Poojitha Thota}, \bibinfo{person}{Krishna~Vamsi Naragam}, {and} \bibinfo{person}{Shirin Nilizadeh}.} \bibinfo{year}{2024}\natexlab{}.
\newblock \showarticletitle{From Chatbots to Phishbots?: Phishing Scam Generation in Commercial Large Language Models}. In \bibinfo{booktitle}{\emph{2024 IEEE Symposium on Security and Privacy (SP)}}. \bibinfo{pages}{36--54}.
\newblock
\urldef\tempurl%
\url{https://doi.org/10.1109/SP54263.2024.00182}
\showDOI{\tempurl}


\bibitem[Rugangazi and Okeyo(2023)]%
        {b9}
\bibfield{author}{\bibinfo{person}{Belyse Rugangazi} {and} \bibinfo{person}{George Okeyo}.} \bibinfo{year}{2023}\natexlab{}.
\newblock \showarticletitle{Detecting Phishing Attacks Using Feature Importance-Based Machine Learning Approach}. In \bibinfo{booktitle}{\emph{2023 IEEE AFRICON}}. \bibinfo{pages}{1--6}.
\newblock
\urldef\tempurl%
\url{https://doi.org/10.1109/AFRICON55910.2023.10293475}
\showDOI{\tempurl}


\bibitem[Sahingoz et~al\mbox{.}(2019)]%
        {sahingoz2019_ML_for_phishing}
\bibfield{author}{\bibinfo{person}{Ozgur~Koray Sahingoz}, \bibinfo{person}{Ebubekir Buber}, \bibinfo{person}{Onder Demir}, {and} \bibinfo{person}{Banu Diri}.} \bibinfo{year}{2019}\natexlab{}.
\newblock \showarticletitle{Machine learning based phishing detection from URLs}.
\newblock \bibinfo{journal}{\emph{Expert Systems with Applications}}  \bibinfo{volume}{117} (\bibinfo{year}{2019}), \bibinfo{pages}{345--357}.
\newblock


\bibitem[Sarasjati et~al\mbox{.}(2022)]%
        {b7}
\bibfield{author}{\bibinfo{person}{Wendy Sarasjati}, \bibinfo{person}{Supriadi Rustad}, \bibinfo{person}{Purwanto}, \bibinfo{person}{Heru~Agus Santoso}, \bibinfo{person}{Muljono}, \bibinfo{person}{Abdul Syukur}, \bibinfo{person}{Fauzi~Adi Rafrastara}, {and} \bibinfo{person}{De~Rosal Ignatius Moses~Setiadi}.} \bibinfo{year}{2022}\natexlab{}.
\newblock \showarticletitle{Comparative Study of Classification Algorithms for Website Phishing Detection on Multiple Datasets}. In \bibinfo{booktitle}{\emph{2022 International Seminar on Application for Technology of Information and Communication (iSemantic)}}. \bibinfo{pages}{448--452}.
\newblock
\urldef\tempurl%
\url{https://doi.org/10.1109/iSemantic55962.2022.9920475}
\showDOI{\tempurl}


\bibitem[Shibli et~al\mbox{.}(2024)]%
        {abuseGPT_ISDFS2024}
\bibfield{author}{\bibinfo{person}{Ashfak~Md Shibli}, \bibinfo{person}{Mir Mehedi~A. Pritom}, {and} \bibinfo{person}{Maanak Gupta}.} \bibinfo{year}{2024}\natexlab{}.
\newblock \showarticletitle{AbuseGPT: Abuse of Generative AI ChatBots to Create Smishing Campaigns}. In \bibinfo{booktitle}{\emph{2024 12th International Symposium on Digital Forensics and Security (ISDFS)}}. \bibinfo{pages}{1--6}.
\newblock
\urldef\tempurl%
\url{https://doi.org/10.1109/ISDFS60797.2024.10527300}
\showDOI{\tempurl}


\bibitem[Tan et~al\mbox{.}(2018)]%
        {8455975}
\bibfield{author}{\bibinfo{person}{Guolin Tan}, \bibinfo{person}{Peng Zhang}, \bibinfo{person}{Qingyun Liu}, \bibinfo{person}{Xinran Liu}, \bibinfo{person}{Chunge Zhu}, {and} \bibinfo{person}{Fenghu Dou}.} \bibinfo{year}{2018}\natexlab{}.
\newblock \showarticletitle{Adaptive Malicious URL Detection: Learning in the Presence of Concept Drifts}. In \bibinfo{booktitle}{\emph{2018 17th IEEE International Conference On Trust, Security And Privacy In Computing And Communications/ 12th IEEE International Conference On Big Data Science And Engineering (TrustCom/BigDataSE)}}. \bibinfo{pages}{737--743}.
\newblock
\urldef\tempurl%
\url{https://doi.org/10.1109/TrustCom/BigDataSE.2018.00107}
\showDOI{\tempurl}


\bibitem[Thomson(2023)]%
        {phish_threat_graham}
\bibfield{author}{\bibinfo{person}{Graham Thomson}.} \bibinfo{year}{2023}\natexlab{}.
\newblock \bibinfo{booktitle}{\emph{Phishing Outlook 2023: Statistics, Real-Life Incidents, and Best Practices}}.
\newblock \bibinfo{type}{{T}echnical {R}eport}. \bibinfo{address}{Albuquerque, NM, USA}.
\newblock


\bibitem[Verma and Das(2017)]%
        {verma2017s_malicious_url}
\bibfield{author}{\bibinfo{person}{Rakesh Verma} {and} \bibinfo{person}{Avisha Das}.} \bibinfo{year}{2017}\natexlab{}.
\newblock \showarticletitle{What's in a url: Fast feature extraction and malicious url detection}. In \bibinfo{booktitle}{\emph{Proceedings of the 3rd ACM on International Workshop on Security and Privacy Analytics}}. \bibinfo{pages}{55--63}.
\newblock


\bibitem[Vrbančič et~al\mbox{.}(2020)]%
        {b10}
\bibfield{author}{\bibinfo{person}{Grega Vrbančič}, \bibinfo{person}{Iztok Fister}, {and} \bibinfo{person}{Vili Podgorelec}.} \bibinfo{year}{2020}\natexlab{}.
\newblock \showarticletitle{Datasets for phishing websites detection}.
\newblock \bibinfo{journal}{\emph{Data in Brief}}  \bibinfo{volume}{33} (\bibinfo{year}{2020}), \bibinfo{pages}{106438}.
\newblock
\showISSN{2352-3409}
\urldef\tempurl%
\url{https://doi.org/10.1016/j.dib.2020.106438}
\showDOI{\tempurl}


\bibitem[Winson(2024)]%
        {b11}
\bibfield{author}{\bibinfo{person}{Winson}.} \bibinfo{year}{2024}\natexlab{}.
\newblock \bibinfo{title}{Dataset for link phishing detection}.
\newblock \bibinfo{howpublished}{https://www.kaggle.com/datasets/winson13/dataset-for-link-phishing-detection}.
\newblock
\newblock
\shownote{(Accessed \ on \ 15 \ May, \ 2024)}.


\bibitem[Xu et~al\mbox{.}(2013)]%
        {codaspy_2013_xu_cross_layer_malwebsite_detection}
\bibfield{author}{\bibinfo{person}{Li Xu}, \bibinfo{person}{Zhenxin Zhan}, \bibinfo{person}{Shouhuai Xu}, {and} \bibinfo{person}{Keying Ye}.} \bibinfo{year}{2013}\natexlab{}.
\newblock \showarticletitle{Cross-layer detection of malicious websites}. In \bibinfo{booktitle}{\emph{Proceedings of the Third ACM Conference on Data and Application Security and Privacy}} (San Antonio, Texas, USA) \emph{(\bibinfo{series}{CODASPY '13})}. \bibinfo{publisher}{Association for Computing Machinery}, \bibinfo{address}{New York, NY, USA}, \bibinfo{pages}{141–152}.
\newblock
\showISBNx{9781450318907}
\urldef\tempurl%
\url{https://doi.org/10.1145/2435349.2435366}
\showDOI{\tempurl}


\bibitem[Yilmaz et~al\mbox{.}(2020)]%
        {denis_mal_domain_2020}
\bibfield{author}{\bibinfo{person}{Ibrahim Yilmaz}, \bibinfo{person}{Ambareen Siraj}, {and} \bibinfo{person}{Denis Ulybyshev}.} \bibinfo{year}{2020}\natexlab{}.
\newblock \showarticletitle{Improving DGA-Based Malicious Domain Classifiers for Malware Defense with Adversarial Machine Learning}. In \bibinfo{booktitle}{\emph{2020 IEEE 4th Conference on Information \& Communication Technology (CICT)}}. \bibinfo{pages}{1--6}.
\newblock
\urldef\tempurl%
\url{https://doi.org/10.1109/CICT51604.2020.9311925}
\showDOI{\tempurl}


\bibitem[Zamir et~al\mbox{.}(2020)]%
        {zamir2020phishing}
\bibfield{author}{\bibinfo{person}{Ammara Zamir}, \bibinfo{person}{Hikmat~Ullah Khan}, \bibinfo{person}{Tassawar Iqbal}, \bibinfo{person}{Nazish Yousaf}, \bibinfo{person}{Farah Aslam}, \bibinfo{person}{Almas Anjum}, {and} \bibinfo{person}{Maryam Hamdani}.} \bibinfo{year}{2020}\natexlab{}.
\newblock \showarticletitle{Phishing web site detection using diverse machine learning algorithms}.
\newblock \bibinfo{journal}{\emph{The Electronic Library}} \bibinfo{volume}{38}, \bibinfo{number}{1} (\bibinfo{year}{2020}), \bibinfo{pages}{65--80}.
\newblock


\bibitem[Zeng et~al\mbox{.}(2020)]%
        {zeng2020diverse}
\bibfield{author}{\bibinfo{person}{Victor Zeng}, \bibinfo{person}{Shahryar Baki}, \bibinfo{person}{Ayman~El Aassal}, \bibinfo{person}{Rakesh Verma}, \bibinfo{person}{Luis Felipe~Teixeira De~Moraes}, {and} \bibinfo{person}{Avisha Das}.} \bibinfo{year}{2020}\natexlab{}.
\newblock \showarticletitle{Diverse datasets and a customizable benchmarking framework for phishing}. In \bibinfo{booktitle}{\emph{Proceedings of the Sixth International Workshop on Security and Privacy Analytics}}. \bibinfo{pages}{35--41}.
\newblock


\end{thebibliography}

\appendix

\end{document}